\documentclass[12pt]{article}
\pdfoutput=1
\usepackage{putex}
\usepackage{graphicx}
\usepackage{epstopdf}
\usepackage{xspace}

\def\MyAlpha{a}

\newcommand{\AdS}{\ensuremath{\mbox{AdS}_5}\xspace}

\begin{document}

\preprint{PUPT-2290}

\institution{PU}{Joseph Henry Laboratories, Princeton University, Princeton, NJ 08544, USA}

\title{Off-center collisions in \AdS with applications to multiplicity estimates in heavy-ion collisions}

\authors{Steven S. Gubser,\footnote{e-mail: {\tt ssgubser@Princeton.EDU}}
Silviu S. Pufu,\footnote{e-mail: {\tt spufu@Princeton.EDU}} and
Amos Yarom\footnote{e-mail: {\tt ayarom@Princeton.EDU}}}

\abstract{We study the trapped surface produced by an off-center collision of light-like, point-sourced shock waves in anti-de Sitter space.  We find an analytic expression for the shape of the trapped surface in the limit where the energy of the shock waves is large and the impact parameter is not too large.  We use the area of the trapped surface to estimate a lower bound on the entropy produced in the collision.  We compare our results to particle multiplicity measurements in heavy-ion collisions as interpreted through the Glauber model.  In an attempt to roughly simulate the effects of asymptotic freedom and confinement in quantum chromodynamics, we also consider the effects of slicing off parts of anti-de Sitter space.}

\date{February 2009}

\maketitle
\tableofcontents

\section{Introduction}
\label{INTRODUCTION}

The Relativistic Heavy Ion Collider (RHIC) collides gold nuclei at $\sqrt{s_{NN}} = 200\,{\rm GeV}$.  This means that each nucleus has energy $E = 100\,{\rm GeV}$ per nucleon, for a total of about $19.7\,{\rm TeV}$ for each nucleus.  The total number  of charged particles $N_{\rm charged}$ that emerge from such a collision can be as large as $5000$: see for example \cite{Back:2004je}.  In \cite{Gubser:2008pc} we pointed out that this number can be approximately reproduced starting from collisions of gravitational shock waves \cite{Aichelburg:1970dh, Dray:1984ha, Hotta:1992qy, Sfetsos:1994xa, Podolsky:1997ni, Horowitz:1999gf, Emparan:2001ce, Arcioni:2001my, Kang:2004jd, Cornalba:2006xk} in $\AdS$, following the methods of the gauge-string duality \cite{Maldacena:1997re,Gubser:1998bc,Witten:1998qj} and identifying the total energy of each nucleus with the energy of the corresponding shock wave.  The calculations of \cite{Gubser:2008pc} rely upon finding a marginally trapped surface in the five-dimensional geometry and using its area to put a lower bound on the entropy of the black hole produced from the collision, following the general plan of \cite{Penrose,DEath:1992hb,DEath:1992hd,DEath:1992qu}.  In \cite{Gubser:2008pc}, we translated this entropy bound into an approximate lower bound on $N_{\rm charged}$ using a fairly well established relation,\footnote{The simplest justification for this relation is that it holds, approximately, for a thermally equilibrated gas of non-interacting hadrons at temperatures just below the transition temperature $T_c \approx 170\,{\rm MeV}$ of QCD \cite{Sollfrank:1992ru,Nonaka:2005vr,Muller:2005en}.}
 \eqn{SN}{
  S \approx 7.5 N_{\rm charged} \,,
 }
between $N_{\rm charged}$ and the entropy $S$ produced in a heavy-ion collision \cite{Sollfrank:1992ru,Pal:2003rz,Nonaka:2005vr,Muller:2005en,Cleymans:2005km}.  A virtue of the trapped surface calculation is that there are no free parameters: the gauge coupling doesn't enter to leading order in a strong coupling expansion, and the overall normalization of the entropy is fixed by the equation of state for static plasmas (see for example \cite{Karsch:2001cy}).  On the downside, the predicted lower bound on $N_{\rm charged}$ scales as $s_{NN}^{1/3}$, which is a faster energy dependence than the $s_{NN}^{1/4}$ scaling predicted by the Landau model \cite{Landau:1953gs} and largely obeyed by the data.
Since the trapped surface computation gives a lower bound on the entropy produced, there is no conflict between \cite{Gubser:2008pc} and experiment---so far.  Conflict will arise if the growth of $N_{\rm charged}$ remains slower than $s_{NN}^{1/3}$ significantly above $\sqrt{s_{NN}} = 200\,{\rm GeV}$.

In this paper we have two main aims.  First, we want to generalize the methods of \cite{Gubser:2008pc} to handle off-center collisions of gravitational shock waves in $\AdS$.  Our generalization amounts to giving an approximate answer to a well-posed question: What is the area of the marginally trapped surface lying on the past light-cone of the shock waves?  The answer, for trapped surfaces that are significantly bigger\footnote{More precisely, a sufficient condition for equation \eqref{Atrapped} to be a good approximation is that $\zeta \equiv  \left( 2 E z_* \frac{G_5}{L^3} \right)^{1/3} \gg 1$ and $\beta \ll \zeta$.} than the radius of $\AdS$, is
 \eqn{Atrapped}{
  A_{\rm trapped} = 4\pi G_5 \left( {4E^2 z_*^2 \over G_5/L^3}
    \right)^{1/3} {\sinh^{-1} \beta \over \beta \sqrt{1+\beta^2}} \,,
 }
where $z_*$ is a characteristic transverse length scale, and $\beta = b/2z_*$ is an $\AdS$ version of the impact parameter.  A more general version of \eno{Atrapped} appears as \eno{GotStrapped} (see also \eqref{StrappedBetter}) and represents our main analytic result.

With \eno{Atrapped} in hand, the standard entropy estimate for the black hole created in the collision is
 \eqn{Sbound}{
  S \geq S_{\rm trapped} \equiv {A_{\rm trapped} \over 4G_5} = 
    \left( {4E^2 z_*^2 \over G_5/L^3}
    \right)^{1/3} {\sinh^{-1} \beta \over \beta \sqrt{1+\beta^2}}
     \,.
 }
Using \eno{SN} and fixing parameters in the same way as in \cite{Gubser:2008pc}, we can obtain approximate lower bounds on $N_{\rm charged}$ and compare them with heavy-ion data---or, more precisely, to heavy-ion data as interpreted using the Glauber model.  We find that our bound on $N_{\rm charged}$ has a significantly weaker dependence on the impact parameter than what data plus Glauber indicates for quantum chromodynamics (QCD).  This result, coupled with the $s_{NN}^{1/3}$ dependence of the bound on $N_{\rm charged}$ from trapped surface calculations, may indicate that the agreement found in \cite{Gubser:2008pc} was to some extent fortuitous---or that the best motivated mapping between quantities in QCD and quantities in a strongly coupled field theory is more subtle than we proposed in \cite{Gubser:2008pc}.  In fact, we find a substantially improved fit to the data by identifying the energy of each shock wave with the fraction of the energy of the nucleus carried by nucleons that participate in the collision.\footnote{We thank B.~Cole, J.~Noronha, P.~Steinberg, and B. Zajc for suggesting this approach.}  Figure~\ref{F:ratiosimple} shows how our total multiplicity estimates compare with data from PHOBOS \cite{Back:2005hs} with the original identification of energy as the total energy of the nucleus.  Figure~\ref{F:rescaled} shows the improved fit obtained from the rescaled energy.

Our second aim is to inquire how our calculations might change if we took into account the non-conformal nature of QCD\@.  It's probably because QCD confines that we have to exclude the energy of the non-participating nucleons before we get good agreement between AdS/CFT calculations and total multiplicity data.  We are led to ask, is there a less contrived way of including the effects of confinement? Low-energy processes in QCD should not contribute as much entropy as in a conformal theory: the reason is that the number of degrees of freedom decreases dramatically below the confinement transition.  To incorporate this in our trapped surface calculation, we discard the part of the trapped surface below a certain fixed depth, corresponding to an infrared cutoff.  Also, for sufficiently high energies, asymptotic freedom dictates that interactions become weaker and weaker.  So it seems safe to say that there is not as much entropy production from hard processes in QCD as there is in a strongly coupled conformal theory.  Correspondingly, we slice away the part of the trapped surface which is above a fixed depth, corresponding to an ultraviolet cutoff.  In summary: we go back to identifying the energy of the shock wave with the total energy of the nucleus; we find the trapped surface in pure $\AdS$ without any cutoffs; and finally, we slice away both the ultraviolet and infrared parts of the trapped surface before converting its area to an estimate of the entropy produced in the collision.

This slicing approach is distinct from a hard-wall construction of a holographic dual of QCD in that the shape of the trapped surface doesn't respond in any way to the cutoffs.  This approach is admittedly naive.  Our main defense of it is to note that it is still more naive to neglect violations of conformal invariance altogether.  In a proper treatment, we should replace the infrared cutoff with a holographic renormalization group flow to a confining theory whose equation of state matches that of QCD: see for example \cite{Gubser:2008ny,Gursoy:2008bu} for work along these lines.  The ultraviolet cutoff should in principle be replaced with some hybrid description where Einstein gravity rolls over into perturbative quantum field theory as one approaches the boundary of \AdS.  

The slicing approach described in the previous paragraph leads to a $s_{NN}^{1/6}$ scaling of the lower bound on $N_{\rm charged}$ at large energies.  Heavy ion collisions at the LHC may be able to probe this scaling.

Our analysis is hardly the only attempt at relating heavy-ion collisions to black hole formation from collisions of gravitational shock waves in \AdS.  Other work along these lines includes \cite{Kang:2004jd,Kang:2005bj,Lin:2007fa,Grumiller:2008va,Gubser:2008pc,Albacete:2008vs,Nastase:2008hw} and references therein.  One hazard of our calculation is that black hole formation is complicated, and it's not clear whether the inequality $S \geq S_{\rm trapped}$ is close to being saturated.  For collisions in flat space, the recent study \cite{Sperhake:2008ga} suggests that it is not: instead one has $S \gtrsim 1.5\,S_{\rm trapped}$, where the inequality is approximately saturated in the limit of ultra-relativistic collisions, and $S_{\rm trapped}$ is based on the canonical choice of trapped surface, analogous to the one we use.  Another hazard is that heavy-ion collisions are a multi-scale process involving both perturbative and non-perturbative dynamics.  A supergravity treatment of an $\AdS$ dual can at best give us a reliable handle only on the range of energy scales where QCD is deconfined but still strongly coupled.  Slicing away parts of $\AdS$ is a poor man's approach to dealing with the non-conformal nature of QCD.

The rest of this paper is organized as follows.  In section~\ref{SETUP} we describe the shock wave solutions we need and exhibit a helpful $O(2)$ symmetry of the collision.  In section~\ref{S:Trapped} we find the shape of the trapped surface in a limit where the size of the surface is much larger than the radius of $\AdS$.  In section \ref{SYMPheno} we use the optical Glauber method to relate our gravity computations to the experimental results of \cite{Back:2006yw}.  In section~\ref{SLICE} we describe in detail our approach of slicing off the UV and IR parts of the trapped surface to obtain a revised estimate of $N_{\rm charged}$ that accounts, albeit crudely, for the non-conformal behavior of QCD\@.  We end with a discussion of our results in section~\ref{DISCUSSION}.

When this paper was nearing completion, we received \cite{Lin:2009pn}, which has some overlap with our results.  We will comment on the relation between our results and the ones in \cite{Lin:2009pn} at the end of section~\ref{S:Trapped}.  We also received \cite{Albacete:2009ji}, which takes a somewhat different approach to shock wave collisions in $\AdS$, following the earlier work \cite{Albacete:2008vs}.

\section{Shock waves from point particles}
\label{SETUP}

Point-like massless particles moving in \AdS generate gravitational shock waves: they are sources for the geometry in the sense that the five-dimensional stress-energy tensor has delta-function support on null geodesics.  From the boundary point of view, a collision of two such particles will correspond to a collision of localized energetic objects which we take to be nuclei.  In section~\ref{SINGLE} we discuss the shock wave geometries before the collision and explain their field theory interpretation.  Most of the material in this section is more thoroughly reviewed in \cite{Gubser:2008pc}, where references to the original literature are also given.  In section~\ref{SYMMETRY} we introduce coordinate systems that make the symmetries of the problem manifest.  We show, among other things, how conformal symmetry relates off-center collisions to collisions of objects of unequal size.

\subsection{Shock wave geometries}
\label{SINGLE}

Particles in \AdS can be described by the action
 \eqn{AdSPlusParticle}{
  S = {1 \over 16\pi G_5} \int d^5 x \, \sqrt{g}
    \left[ R + {12 \over L^2} \right] +
    \int d\eta \left[ {1 \over 2e} g_{\mu\nu}
     {dx_*^\mu \over d\eta} {dx_*^\nu \over d\eta} -
     {e \over 2} m^2 \right] \,,
 }
where $x^\mu = x_*^\mu(\eta)$ is the trajectory of the particle, $m$ is its mass, $\eta$ is an arbitrary parametrization of the worldline, and $e$ is the worldline one-bein.  We of course set $m=0$.   Ignoring the back-reaction of the particle, the geometry is pure \AdS:
 \eqn{PureAdS}{
  ds_{AdS_5}^2 = {L^2 \over z^2} \left[
   -(dx^0)^2 + (dx^1)^2 + (dx^2)^2 + (dx^3)^2 + dz^2 \right] \,,
 }
and the particle's trajectory is a null geodesic in this geometry.  The momentum conjugate to $x_*^\mu$ is
 \eqn{ConjugateMomentum}{
  p_\mu = {1 \over e} g_{\mu\nu} {dx_*^\nu \over d\eta} \,.
 }
Because of the translation symmetry in the $x^m$ directions, where $m$ runs from $0$ to $3$, the quantities $p_m$ are conserved.  They can be identified as the flat space four-momentum in the gauge theory.  Let's focus on a particle whose trajectory is
 \eqn{ExampleTrajectory}{
  x_*^0 = x_*^3 = t \qquad x_*^1 = x_*^2 = 0 \qquad
   z_* = \hbox{constant} \,.
 }
Then we have
 \eqn{IdentifyMomenta}{
  p_m = (-E,0,0,E) \,,
 }
where $E$ is the energy of the particle.  Introducing the light-cone coordinates
 \eqn{xppm}{
  x^\pm = x^0 \pm x^3 \qquad p_\pm = {1 \over 2} (p_0 \pm p_3) \,,
 }
one finds from \eno{IdentifyMomenta} that $p_- = -E$.  We may choose $\eta=t$ as the worldline parameter.  Then from \eno{ConjugateMomentum} one finds that $e = {1 \over E} {L^2 \over z^2}$.

Now let's add in the back-reaction.  The Einstein equations following from \eno{AdSPlusParticle} are
 \eqn{EEqs}{
  R_{\mu\nu} - {1 \over 2} g_{\mu\nu} R -
    {6 \over L^2} g_{\mu\nu} = 8\pi G_5 J_{\mu\nu} \,,
 }
where
 \eqn{Jmunu}{
  J_{\mu\nu} = \int d\eta \, {e \over \sqrt{-g}}
    \delta^5(x^\mu-x_*^\mu(\eta)) p_\mu p_\nu
 }
is the stress tensor of the particle.  Using light-cone coordinates, the only non-zero component of $J_{\mu\nu}$ is
 \eqn{Jmm}{
  J_{--} = E {z^3 \over L^3} \delta(x^1) \delta(x^2) \delta(z-z_*)
    \delta(x^-) \,.
 }
Plugging the ansatz
 \eqn{ShockAnsatz}{
  ds^2 = ds_{AdS_5}^2 + {L \over z} \Phi(x^1,x^2,z)
    \delta(x^-) (dx^-)^2
 }
into \eno{EEqs}, one finds that the only non-trivial equation is the $--$ equation, which reads
 \eqn{mmEin}{
  \left( \nabla_{H_3}^2 - {3 \over L^2} \right) \Phi =
    -16\pi G_5 E {z^4 \over L^4} \delta(x^1) \delta(x^2)
      \delta(z-z_*) \,,
 }
where
 \eqn{nablaH}{
  \nabla_{H_3}^2 = {z^2 \over L^2} \left[
    \left( {\partial \over \partial x^1} \right)^2 +
    \left( {\partial \over \partial x^2} \right)^2 +
    {\partial^2 \over \partial z^2} \right] -
    {z \over L^2} {\partial \over \partial z}
 }
is the laplacian on $H_3$, whose line element is
 \eqn{dsH}{
  ds_{H_3}^2 = {L^2 \over z^2} \left[ (dx^1)^2 + (dx^2)^2 + dz^2
    \right] \,.
 }
The differential equation \eno{mmEin} is subject to the boundary condition that $\Phi \to 0$ as one approaches the boundary of $H_3$.  This is equivalent to requiring the perturbation of the metric to vanish at the boundary, meaning that the shock wave describes a state in the dual field theory rather than a deformation of the lagrangian.

Reducing the full non-linear Einstein equations in five dimensions to a single linear differential equation on $H_3$ is a drastic simplification.  It is important to realize that no linearized approximation is needed to derive \eno{mmEin}: a solution to it leads to an exact solution of the Einstein equations.  Only, because $\delta(x^-)$ is involved in the metric \eno{ShockAnsatz}, it is a solution in the sense of distributions.

The solution to \eqref{mmEin} is
 \eqn{PhiOfQ}{
  \Phi = {G_5 E z_* \over 8 L^2} {1 \over q^3} \,
    {}_2F_1(3, 5/2; 5; -1/q) \,,
 }
where
 \eqn{ChordalDistance}{
  q = {(x^1)^2 + (x^2)^2 + (z-z_*)^2 \over 4zz_*} \,,
 }
and ${}_2F_1(a,b;c;z)$ is the hypergeometric function.  The particular hypergeometric function in \eno{PhiOfQ} has a closed form expression which is algebraic in $q$, but its explicit form is not very illuminating. The geodesic distance (in $H_3$) from the point $(0,0,z_*)$ to where the point-source is located is $\ell = 2 L \sinh^{-1} \sqrt{q}$. Thus, the solution to \eqref{mmEin} is invariant under $O(3)$. This $O(3)$ is a subgroup of the isometry group of $H_3$, which is $O(3,1)/{\bf Z}_2$.\footnote{$O(3,1)$ has four connected components.  In the hyperboloid coordinates, to be introduced in \eno{ThreeHyperboloid}, the ${\bf Z}_2$ we divide by acts as $X^{-1} \to -X^{-1}$.  Dividing by an additional ${\bf Z}_2$ acting as parity on the remaining coordinates would give $SO(3,1)$.}

One may extract an expectation value for the gauge theory stress tensor dual to the point source by using a standard expression for one-point functions:
 \eqn{Tmunu}{
  \langle T_{--} \rangle =
    {L^2 \over 4\pi G_5} \delta(x^-) \lim_{z \to 0}
      {\Phi(x^1,x^2,z) \over z^3} =
      {2 E z_*^4 \over \pi (x_\perp^2 + z_*^2)^3} \delta(x^-) \,,
 }
where $x_\perp = (x^1,x^2)$ parameterizes the plane transverse to the collision.  All other components of $\langle T_{mn} \rangle$ vanish.  One can straightforwardly check from \eno{Tmunu} that
 \eqn{TotalEagain}{
  \int d^3 x \, \langle T_{--} \rangle = E
   \qquad\hbox{and}\qquad
  {\int d^3 x \, x_\perp^2 \langle T_{--} \rangle \over
   \int d^3 x \, \langle T_{--} \rangle} = z_*^2 \,.
 }
The first equality shows that $E$ is the total energy of one shock wave in the gauge theory, as it is in the gravitational description.  The second equality shows that the energy-weighted root-mean-square size of the distribution \eno{Tmunu} is $z_*$.  This is an elementary example of the relation between depth in \AdS and size in the dual field theory.

Now consider an off-center collision of two shocks, where the trajectories of the point sources take the form
\begin{equation}
\label{E:Poincaretrajectories}
	x^3 = \mp x^0 \qquad
	x^1 = b_\pm \qquad
	x^2 = 0 \qquad
	z = z_\pm \,.
\end{equation}
A straightforward choice would be to set $b_\pm = \pm b/2$ and $z_+ = z_- = z_*$: then we would be colliding an energy distribution like \eno{Tmunu} with another one of the same size going the other way.  But we can just as easily consider $z_+ \neq z_-$, which means that we are colliding energy distributions with different transverse sizes.

Causality dictates that the gravitational shocks generated by these particles cannot affect one another outside the future light-cone of the collision point.   So one must be able to superpose them to obtain an exact (distributional) solution of the Einstein equations which holds provided $x^+<0$ or $x^-<0$:
 \eqn{TwoShocks}{
  ds^2 = ds_{AdS_5}^2 +
   {L \over z} \Phi_-(x^1,x^2,z) \delta(x^-) (dx^-)^2 +
   {L \over z} \Phi_+(x^1,x^2,z) \delta(x^+) (dx^+)^2 \,.
 }
By the same calculations that led to \eno{PhiOfQ}, one finds
 \eqn{E:PhiVal}{
  \Phi_\pm = {G_5 E_\pm z_\pm \over 8L^2}
    {1 \over q_\pm^3} \, {}_2F_1(3,5/2;5;-1/q_\pm) \,.
 }
Here $E_+$ and $E_-$ are the energies of the two shock waves, which we do not assume to be equal.  The quantities $q_\pm$ are the chordal distances between a given point $(x^1,x^2,z)$ and the location $(b_\pm,0,z_\pm)$ of the shocks:
\begin{equation}
\label{E:Chordalgeneral}
	q_\pm = {(x^1 - b_\pm)^2+(x^2)^2+(z-z_\pm)^2 \over 4zz_\pm} \,.
\end{equation}
Passing through the calculations that led to \eno{Tmunu}, one finds that the gauge-theory shock wave moving in the $+x^3$ direction generates a boundary theory stress-energy tensor
 \eqn{Tmmidentical}{
  \langle T_{--} \rangle = {2 E_- z_-^4 \over
    \pi \left[ (x^1 - b_-)^2 + (x^2)^2 + z_-^2 \right]^3}
    \delta(x^-) \,,
 }
while the shock moving in the $-x^3$ direction generates
 \eqn{Tppidentical}{
  \langle T_{++} \rangle = {2 E z_+^4 \over
    \pi \left[ (x^1 - b_+)^2 + (x^2)^2 + z_+^2 \right]^3}
    \delta(x^+) \,.
 }

While it is difficult to see it in the Poincar\'e coordinate system, the two-shock geometry \eqref{TwoShocks} has an $O(2)$ symmetry. In the following subsection we use a different coordinate system which makes this symmetry explicit and will also make it easier to deal with collisions of unequal-sized energy distributions in the boundary theory.

\subsection{Symmetries of the collision}
\label{SYMMETRY}

In \eno{PureAdS} and~\eno{xppm}, we have introduced the Poincar\'e patch coordinates for \AdS and our conventions for light-cone coordinates.  However, the $O(4,2)$ conformal symmetry becomes more transparent if one uses the hyperboloid coordinates $X^M$ in ${\bf R}^{4,2}$, which are subject to the constraint
\begin{equation}
\label{E:hyperboloid}
    -\left(X^{-1}\right)^2
    -\left(X^{0}\right)^2
    +\left(X^{1}\right)^2
    +\left(X^{2}\right)^2
    +\left(X^{3}\right)^2
    +\left(X^{4}\right)^2
    =-L^2.
\end{equation}
By using
 \eqn{E:GlobalToPoincare}{
  X^{-1} &= \frac{z}{2}\left(1+\frac{L^2+\vec{x}^2-(x^0)^2}{z^2}
   \right)  \cr
  X^m &= L\frac{x^m}{z}  \cr
  X^{4} &= \frac{z}{2}\left(-1+\frac{L^2-\vec{x}^2+(x^0)^2}{z^2}
   \right) \,,
 }
we see that the metric inherited on this hypersurface from the standard flat metric on ${\bf R}^{4,2}$ is the same as \eno{PureAdS}.\footnote{It's worth noting that global \AdS is usually thought of as the covering space of the hyperboloid \eno{E:hyperboloid}, which has closed timelike curves.}

The most general collision we will consider is that of two shock waves whose point sources follow the trajectories
\begin{equation}
\label{E:Particle1}
	X^0 = \mp X^3 \qquad
	X^1 = \pm L \beta \cos a \qquad
	X^2 = 0 \qquad
	X^4= \pm L \beta \sin a \qquad
	X^{-1} = L \sqrt{1+\beta^2} \,.
\end{equation}
Here $\beta$ is a dimensionless \AdS version of the impact parameter.  Changing $a$ evidently amounts to a conformal transformation in the gauge theory.  In Poincar\'e coordinates, \eno{E:Particle1} becomes
\begin{equation}
	x^3 = \mp x^0 \qquad
	x^1 = b_\pm \qquad
	x^2 = 0 \qquad
	z = z_\pm \,,
\end{equation}
where
 \eqn{bzDef}{
  b_\pm \equiv \pm {L \beta \cos a \over
    \sqrt{1 + \beta^2} \pm \beta \sin a} \qquad
  z_\pm \equiv {L \over \sqrt{1+\beta^2} \pm \beta \sin a}
    \,.
 }
By choosing $\MyAlpha=0$, we find ourselves colliding objects of equal size; but if $\MyAlpha \neq 0$, the objects have different sizes.  In short: the relative sizes of colliding objects can be changed by a conformal transformation!  Experiment mostly focuses on colliding equal-sized nuclei, but it would be interesting to inquire to what extent the collision of unequal-sized objects (say, gold against copper) would provide experimental tests of the degree to which conformal symmetry is preserved in the dynamics of a heavy-ion collision. The residual $O(2)$ symmetry which we mentioned earlier manifests itself as rotations which leave the quadratic form $(X^1 \sin \MyAlpha - X^4 \cos \MyAlpha)^2 + (X^2)^2$ invariant.

Most of our analysis will take place on $H_3$, which is the intersection of the null surfaces $x^+ = 0$ and $x^- = 0$ in the Poincar\'e patch.  From \eno{E:GlobalToPoincare} we see that $H_3$ is determined by $X^+ = X^- = 0$---almost.  If we impose these two conditions on \eno{E:hyperboloid}, we obtain the equation
 \eqn{ThreeHyperboloid}{
  -(X^{-1})^2 + (X^1)^2 + (X^2)^2 + (X^4)^2 = -L^2 \,,
 }
which describes a two-sheeted, three-dimensional hyperboloid in ${\bf R}^{3,1}$.  But \eno{E:hyperboloid} also implies that $X^{-1}>0$ in the Poincar\'e patch, and this condition selects the upper sheet.  Sometimes $H_3$ is denoted $H_3^+$ to emphasize this restriction, but we will instead use $H_3$ to mean just the upper sheet.  Setting $x^0 = x^3 = 0$ in \eno{E:GlobalToPoincare} leads immediately to a coordinate transformation between hyperboloid coordinates $(X^{-1},X^1,X^2,X^4)$ on $H_3$ and Poincar\'e coordinates $(x^1,x^2,z)$.

To make the residual $O(2)$ symmetry of $H_3$ transparent, we switch to radial coordinates
\begin{subequations}
\label{E:H3coords}
\begin{align}
  X^{-1} &= \sqrt{L^2 + r^2} \\
  X^1 &= r \cos \theta \cos \MyAlpha -
    r \sin \theta \cos \phi \sin \MyAlpha  \\
  X^2 &= r \sin \theta \sin \phi  \\
  X^4 &= r \cos \theta \sin \MyAlpha +
    r \sin \theta \cos \phi \cos \MyAlpha \,,
\end{align}
\end{subequations}
in which case the line element on $H_3$ takes the form
\begin{equation}
\label{E:lineH3r}
	ds_{H_3}^2 = \frac{L^2}{L^2+r^2}dr^2+
	  r^2(d\theta^2+\sin^2\theta d\phi^2) \,.
\end{equation}
In this coordinate system the chordal distance \eqref{E:Chordalgeneral} takes the form
 \eqn{FoundQpm}{
  q_\pm = -{1 \over 2} +
    {\sqrt{L^2+r^2} \sqrt{1+\beta^2} \over 2L} \mp
    {r\beta \cos\theta \over 2L} \,
 }
and the $O(2)$ symmetry of the solution is made clear by the invariance of \eqref{E:PhiVal} under rotations of $\phi$.  (Recall that setting $a=0$ corresponds in gauge theory to colliding objects of equal size, separated in the $x^1$ direction.)

\section{Trapped surface computation}
\label{S:Trapped}

If the impact parameter is not too large, then after the collision a black hole will probably form. To obtain a lower bound on the entropy of the black hole produced in such a collision, we use the method of \cite{Penrose}, which was further developed in \cite{DEath:1992hb,DEath:1992hd,DEath:1992qu,Eardley:2002re} and adapted to AdS space in \cite{Hotta:1992qy,Sfetsos:1994xa,Podolsky:1997ni,Emparan:2001ce,Arcioni:2001my,Nastase:2004pc,Gubser:2008pc,AlvarezGaume:2008fx}.  The method is to find a marginally trapped surface ${\cal S}$, composed of two parts: ${\cal S} = {\cal S}_+ \cup {\cal S}_-$ where the ${\cal S}_i$'s are parameterized by
\begin{equation}
{\cal S}_{\pm} = \left\{x^\mu \in \AdS:\ x^\mp=0,\,x^\pm=-\Psi_\pm(x^1,x^2,z)\right\} \,,
\end{equation}
and then use the area theorem and the Cosmic Censorship conjecture to give a lower bound on the entropy produced:
\begin{equation}
\label{E:AtoSbound}
	S \geq S_{\rm trapped} \equiv \frac{A_{\rm trapped}}{4 G_5} \,,
\end{equation}
where $A_{\rm trapped}$ is the area of ${\cal S}$.
The condition that ${\cal S}$ is a marginally trapped surface (meaning that it has zero expansion) can be translated into an unusual boundary problem: the functions $\Psi_{\pm}$ need to satisfy
\begin{subequations}
\label{E:EOM}
\begin{equation}
\label{E:kinetic}
    \left(\nabla_{H_3}^2-\frac{3}{L^2}\right)\left(\Psi_{\pm}-\Phi_{\pm}\right)=0
\end{equation}
and
\begin{align}
\label{E:DirichletBC}
    \Psi_{\pm}\Big|_{\mathcal{C}}&=0\\
\label{E:NeumannBC}
    g^{ab} \partial_a \Psi_+ \partial_b \Psi_-\Big|_{\mathcal{C}} &= 4
\end{align}
\end{subequations}
where $\mathcal{C}$ is the curve (within the collision plane $x^+ = x^- = 0$) on which the surfaces ${\cal S}_+$ and ${\cal S}_-$ intersect.\footnote{As before, we use the words ``curve'' and ``surface'' despite the fact that ${\cal C}$ has two dimensions and ${\cal S}$ has three.}  The indices $a$, $b$ run over the $H_3$ direction, and $g^{ab}$ is the inverse of the metric \eno{E:lineH3r}.

Solving \eqref{E:EOM} exactly is difficult when the collision is off-center.  However, an appropriate perturbative expansion leads to an analytically tractable problem.
Consider the parameters
 \eqn{ZetaPM}{
  \zeta_\pm = \left( 2 E_\pm z_\pm {G_5 \over L^3} \right)^{1/3}
    \,,
 }
and define\footnote{For comparison with \cite{Gubser:2008pc}, it is useful to note that for central collisions in \AdS with $z_+ = z_- = L$, $\zeta^3 = 4 q (1+q)(1+2q)$ and $x = 2 \sqrt{q(1+q)}$.  So when $\zeta$ is large, $\zeta \approx 2q \approx x$.} 
 \eqn{ZetaDef}{
  \zeta \equiv  \sqrt{\zeta_+ \zeta_-} \,.
 }
By boosting in the $x^3$ direction, we can change the ratio $E_+/E_-$ without changing $z_+$ or $z_-$.  In particular, we can choose a boost parameter such that $E_+ \to \lambda E_+$ and $E_- \to \lambda^{-1} E_-$ with $\lambda = \sqrt{E_-/E_+}$.  After such a boost, we are in a frame where $\zeta_+ = \zeta_- = \zeta$.  We will generally prefer to work in such a frame.

In \cite{Gubser:2008pc} it was noticed that in the case of central collisions of identical objects, i.e., $\beta=\MyAlpha=0$, the size of the trapped surface grows linearly in $\zeta$ when $\zeta$ is large.  It is plausible that the same is true of non-central collisions, at least when the impact parameter $\beta$ is held fixed while $\zeta$ is made large.  Let's define a new radial coordinate:
 \eqn{E:Defrho}{
  \rho \equiv {r \over \zeta L} \,.
 }
The metric on $H_3$ takes the form
 \eqn{zetaMet}{
  ds_{H_3}^2 = L^2 \left[ {d\rho^2 \over \rho^2 + 1/\zeta^2} +
    \zeta^2 \rho^2 (d\theta^2 + \sin^2 \theta d\phi^2) \right] \,.
 }
We have already remarked that the off-center collision respects the $O(2)$ symmetry generated by additive shifts of $\phi$.  So the trapped surface must be a surface of revolution in the $\phi$ direction.  The curve ${\cal C}$ must likewise have the $O(2)$ symmetry, which means that its position in $H_3$ can be parameterized as $\rho = \rho_{\cal C}(\theta)$.  Because we have chosen to work in a frame where $E_+ z_+ = E_- z_-$, there is an additional ${\bf Z}_2$ reflection symmetry which interchanges the shocks by sending $\theta \to \pi-\theta$ and at the same time $x^3 \to -x^3$.  As consequences of this symmetry, we have
 \eqn{ReflectRelate}{
  \Psi_-(\rho,\theta) = \Psi_+(\rho,\pi-\theta) \qquad\qquad
    \rho_{\cal C}(\theta) = \rho_{\cal C}(\pi-\theta) \,.
 }
If we define
 \eqn{hpmDefine}{
  L h_\pm(\rho,\theta) =
   \Psi_\pm(\rho,\theta) - \Phi_\pm(\rho,\theta)
    \,,
 }
then the functions $h_\pm$ are eigenfunctions of the laplacian on $H_3$:
\begin{subequations}\label{RecastDef}
 \eqn{hpmEOM}{
  \left( \nabla_{H_3}^2 - {3 \over L^2} \right) h_\pm =
    {1 \over L^2} \left[ \rho^2 \partial_\rho^2 +
      3 \rho \partial_\rho - 3 +
     {1 \over \zeta^2 \rho^2} \left(
       {1 \over \sin\theta} \partial_\theta \sin\theta
        \partial_\theta + \rho^2 \partial_\rho +
        2\rho \partial_\rho \right) \right] h_\pm = 0 \,.
 }
The boundary conditions \eno{E:DirichletBC} and \eno{E:NeumannBC} can be re-expressed as
 \begin{align}
  \bigg[ h_\pm + {\Phi_\pm \over L} \bigg]_{\cal C} &= 0
     \label{E:Dagain1} \\
  \Bigg[ \left( \rho^2 + {1 \over \zeta^2} \right)
    \partial_\rho \left( h_+ + {\Phi_+ \over L} \right)
    \partial_\rho \left( h_- + {\Phi_- \over L} \right)
     \qquad\qquad &  \nonumber \\ +
    {1 \over \zeta^2 \rho^2}
    \partial_\theta \left( h_+ + {\Phi_+ \over L} \right)
    \partial_\theta \left( h_- + {\Phi_- \over L} \right)
    \Bigg]_{\cal C} &= 4 \,.
\label{E:Dagain2}
\end{align}
\end{subequations}
The functions $h_\pm$ obey the same symmetry relation as $\Psi_\pm$: $h_-(\rho,\theta) = h_+(\rho,\pi-\theta)$.

To make \eqref{RecastDef} analytically tractable, we expand
 \eqn{hRhoExpand}{
  h_+(\rho,\theta) &= h_0(\rho,\theta) + {1 \over \zeta^2}
    h_2(\rho,\theta) +
    \ldots  \cr
  \rho_C(\theta) &= \rho_0(\theta) + {1 \over \zeta^2} \rho_2(\theta)
     + \ldots \,.
 }
The differential equation \eno{hpmEOM} can now be broken down order-by-order in $\zeta$.
At leading order in $\zeta$ we find
\eqn{hpmEOMordered}{
  \left( \rho^2 \partial_\rho^2 + 3 \rho \partial_\rho - 3
     \right) h_0 &= 0\,.
 }
The solution to \eqref{hpmEOMordered} is
\begin{equation}
\label{hZeropresol}
    h_0(\rho,\theta) = C_0(\theta)\rho + D_0(\theta)\rho^{-3}.
\end{equation}
The second term on the right hand side of \eno{hZeropresol} can be discarded.  Naively, the reason is that it is singular at $\rho=0$.  This is an unsatisfactory argument because $\zeta\rho = r/L$, so that working in the large $\zeta$ regime implies that $r$ must be large.  In order to impose the correct boundary conditions at small $\rho$ we need, for example, to match the small $\rho$ behavior of \eqref{hZeropresol} to the large $r$ asymptotics of the full solution to \eqref{RecastDef}. We do this in appendix~\ref{MATCH} where we find that the naive expectation $D_0(\theta)=0$ holds. Thus,
\begin{equation}
  h_0 = C_0(\theta) \rho  \label{hZeroSoln}.
\end{equation}

At subleading order in $\zeta$ we have
\eqn{hpmEOMordered2}{
  \left( \rho^2 \partial_\rho^2 + 3 \rho \partial_\rho - 3
     \right) h_2 &= -{1 \over \rho^2}
      \left( \rho^2 \partial_\rho^2 + 2 \rho \partial_\rho +
       \partial_\theta^2 + \cot\theta \partial_\theta \right) h_0  \,.
}
The solution to \eqref{hpmEOMordered2} is
\begin{equation}
  h_2 = C_2(\theta) \rho + {1 \over 4\rho} \left( \partial_\theta^2 +
    \cot\theta \partial_\theta + 2 \right) C_0(\theta) \,.
    \label{hTwoSoln}
\end{equation}
As was the case for $h_0$, the solution to \eqref{hpmEOMordered2} may also include a $D_2(\theta) \rho^{-3}$ term. In appendix \ref{MATCH} we also show that $D_2(\theta) = 0$. One should be able, in principle, to carry out this procedure to arbitrary order in $\zeta$. We find that at relative order $\zeta^{-4}$ some of the homogeneous solutions proportional to $\rho^{-3}$ (which we may call $D_4(\theta)$ following the notation in the preceding paragraphs) may not vanish. Additionally, there will also be a contribution at order $\zeta^{-4} \log \zeta$.   So going beyond the order shown explicitly in \eno{hTwoSoln} seems to present some new difficulties.  Fortunately, for practical purposes, $\zeta$ is numerically fairly large: $\zeta \sim 50$ or more in the cases we'll be considering.  So we will not concern ourselves further with higher order corrections, and work only through relative order $1/\zeta^2$.

To determine the remaining integration constant $C_i(\theta)$, we need to implement the boundary conditions \eqref{E:Dagain1} and \eqref{E:Dagain2}. At leading order in $\zeta$, \eqref{E:Dagain1} reads
\begin{equation}
\label{E:DiricheltO0}
	\rho_0 C_0(\theta) + \frac{1}{2 \rho_0^3 \left(\sqrt{1+\beta^2}-\beta \cos\theta\right)^3} = 0 \,.
\end{equation}
Plugging this into \eqref{E:Dagain2} we get
\begin{equation}
	\rho_0(\theta) = \frac{1}{\sqrt{1+\beta^2\sin^2\theta}} \,.
\end{equation}
This is the leading order expression for $\rho_{\mathcal{C}}$ when $\zeta \gg 1$.
At subleading order we find:
\begin{equation}
\label{E:rhofinal}
	\rho_{\mathcal{C}}(\theta)=\frac{1}{\sqrt{1+\beta^2\sin^2\theta}}
	+
	\frac{-1+\beta ^2 (4 \cos (2 \theta )-3)-6 \beta ^4 \sin ^2\theta}{6 \zeta^2 (1+\beta^2\sin^2\theta)^{3/2}}
    + {\cal O} (\zeta^{-4})\,.
\end{equation}

Reading off the volume element on $H_3$ from the metric \eqref{zetaMet}, one can compute the area of the trapped surface ${\cal S}$ from
 \eqn{ATrappedDef}{
  A_{\rm trapped} = 2 \int_0^{2 \pi} d \phi \int_0^\pi d\theta \int_0^{\rho_C(\theta)} d\rho\,
    {L^3 \zeta^2 \rho^2 \sin\theta \over \sqrt{\rho^2 + 1/\zeta^2}} \,.
 }
The explicit factor of $2$ that multiplies the integral in \eqref{ATrappedDef} comes from the fact that ${\cal S}$ consists of two parts, ${\cal S}_+$ and ${\cal S}_-$, each with an area equal to the integral of the $H_3$ volume element over the region $\rho \leq \rho_C(\theta)$.  Plugging \eqref{E:rhofinal} into \eqref{ATrappedDef} and using \eqref{E:AtoSbound}, we find that to leading order at large $\zeta$ we have
\begin{equation} \label{GotStrapped}
	S \geq S_{\rm trapped} =
		\left(\frac{4 E_+ E_- z_+ z_-}{G_5/L^3}\right)^{1/3}	\pi	\frac{\sinh^{-1} \beta}{\beta \sqrt{1+\beta^2}} \,.
\end{equation}
An expression for $\beta$ in terms of the impact parameter $b = b_+ - b_-$ and the rms sizes $z_+$ and $z_-$ of the flattened energy distributions in the boundary theory (see \eqref{Tmmidentical} and \eqref{Tppidentical}) can be obtained from \eqref{bzDef}:
\begin{equation}
\label{E:betatobz}
	\beta =  \frac{1}{2}\sqrt{\frac{b^2+(z_+-z_-)^2}{z_+z_-}} \,.
\end{equation}
Plugging this expression into \eqref{GotStrapped}, one can obtain an expression for $S_{\rm trapped}$ only in terms of quantities defined in the boundary CFT\@.  The special case of \eno{GotStrapped} where $z_+ = z_- = z_*$ and $E_+ = E_- = E$ is equivalent to the result \eno{Atrapped} that we quoted in the introduction.

Subleading corrections to \eqref{GotStrapped} can be straightforwardly worked out from equations \eqref{E:rhofinal} and \eqref{ATrappedDef}.  One obtains
 \eqn{StrappedBetter}{
  S_{\rm trapped} = \pi {L^3 \over G_5} \left[
    {\sinh^{-1} \beta \over \beta \sqrt{1 + \beta^2}} \zeta^2
    - \log \zeta 
    - \left(\log 2 - {2 \over 3} + 
    {(1 + 2 \beta^2) \sinh^{-1} \beta \over 2 \beta \sqrt{1 + \beta}} \right) 
    + {\cal O}(1/\zeta^2) \right] \,,
 }
where $\zeta$ is as defined in \eqref{ZetaDef}.  The non-analytic $\log \zeta$ term in the above expression comes from the fact that $1/\zeta^2$ acts as a small $\rho$ regulator in the area integral \eqref{ATrappedDef}.  We can estimate that \eqref{StrappedBetter} will break down when $\beta \sim \zeta$, because then the constant term in \eqref{StrappedBetter} becomes of the same order as the leading term, so one would expect that higher order terms in the series will then also be important.    If, instead of colliding point-sourced shock waves in $\AdS$, we collide shock waves whose sources are spread out in the transverse plane, then we expect that the trapped surface will be almost unaffected provided it extends over a much bigger region of $H_3$ than the sources do.  There is a precise result along these lines for head-on collisions \cite{Gubser:2008pc}; see also the related discussion \cite{AlvarezGaume:2008fx}.

\begin{figure}
  \centerline{\includegraphics[width=5in]{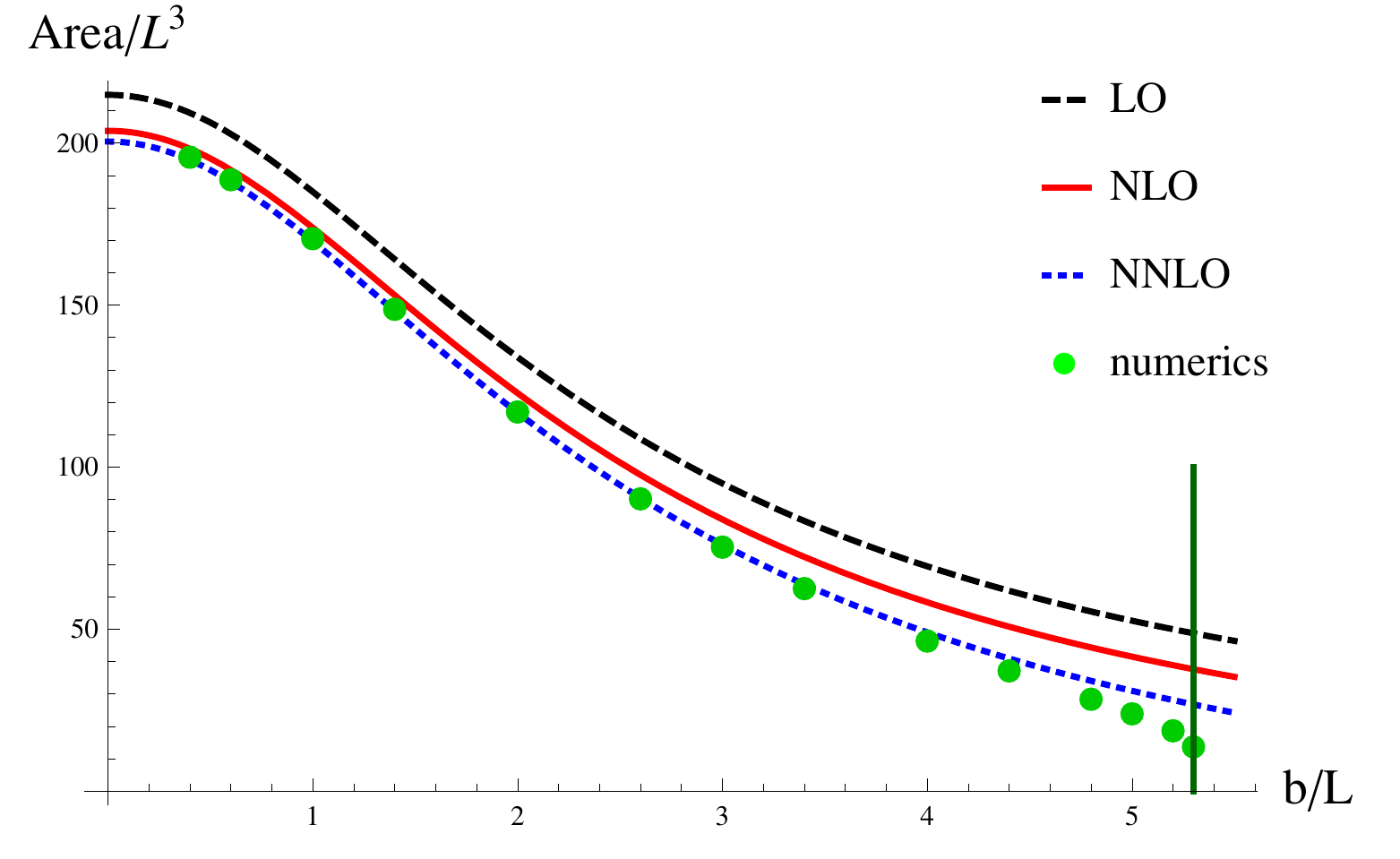}}
  \caption{(Color online.)  Comparisons between the numerics of \cite{Lin:2009pn} and the analytic formula \eqref{StrappedBetter}.  The black dashed curve represents the leading term in \eqref{StrappedBetter};  the solid red curve corresponds to the first two terms in \eqref{StrappedBetter}; the dotted blue curve represents the expression \eqref{StrappedBetter}, which is correct up to a term of order ${\cal O}(1/\zeta^2)$; the green dots represent the numerical evaluations used in figure~3 of \cite{Lin:2009pn};  lastly, the vertical green line marks the place where, according to \cite{Lin:2009pn}, the maximum impact parameter $b_{\rm max} / L$ occurs.  We thank S.~Lin and E.~Shuryak for providing us with the results of their numerical evaluations.}\label{slComparison}
 \end{figure}

The area of the trapped surface was computed numerically in \cite{Lin:2009pn} for particular values of $\zeta$, with $z_* = L$.    In figure~\ref{slComparison} we compare the numerical results of \cite{Lin:2009pn} (in the case where $G_5 E / L^2 = 100$, corresponding to $\zeta \approx 5.848$) to the analytical prediction \eqref{StrappedBetter}.    As can be seen from this figure, equation \eqref{StrappedBetter} is a good approximation whenever $b/L \lesssim 4$ (corresponding to $\beta \lesssim 2$), and breaks down for larger values of $b$.  In particular, the approximate bound \eqref{StrappedBetter} doesn't capture the fact that there exists a maximum value $b = b_{\rm max}$ above which marginally trapped surfaces of the type considered above no longer exist.  This doesn't necessarily imply that in collisions with impact parameters larger than $b_{\rm max}$ there is no black hole formation, since there could be marginally trapped surfaces elsewhere.  The existence of $b_{\rm max}$ is however suggestive of an upper limit on the impact parameter for black hole formation.  

The values $\zeta \lesssim 6$ used in \cite{Lin:2009pn} are smaller than the range we think can be compared most naturally to heavy-ion collisions in the range of energies attained at RHIC.  As we shall explain in section~\ref{SYMPheno}, we prefer values of $\zeta$ more than ten times bigger.  For such large $\zeta$, even the first correction to \eno{GotStrapped} is insignificant for the range of $b/L$ we will be interested in, namely $b/L$ no more than a few.

As one approaches $b_{\rm max}$, the flat space analysis of shock-wave collisions described in \cite{Yoshino:2002tx} suggests that the slope of $S_{\rm trapped}(b)$ becomes infinite.  This type of behavior implies that at $b = b_{\rm max}$ subleading terms in the series expansion \eqref{StrappedBetter} are important, and by the previous discussion, the maximum impact parameter leading to black hole formation satisfies $b_{\rm max} \gtrsim \zeta L$.  Because $\zeta \propto E^{1/3}$, a natural expectation is $b_{\rm max} \sim E^{1/3}$ for large energies.  This scaling seems consistent with the observation of \cite{Lin:2009pn} that for $\zeta \lesssim 6$ one has $b_{\rm max} \sim E^\alpha$ with $\alpha \approx 0.37$.  Because $b_{\rm max} / \zeta L \approx 0.91$ according to \cite{Lin:2009pn} for $\zeta = 5.848$, and because the difference between the scaling with energy found there and the one we expect is fairly small, it is plausible that the estimate $b_{\rm max} \approx \zeta L$ is within a factor of $2$ of the correct answer for $\zeta$ larger than a few.

\section{Entropy production in a strongly coupled conformal field theory}
\label{SYMPheno}

The motivation for our analysis is to gain some insight into entropy production in heavy-ion collisions.  This implies that we need to somehow translate our \AdS trapped surface results into expectations for QCD\@.  Any such translation is perilous, because what gravity calculations in $\AdS$ are really dual to is a strongly coupled conformal field theory, and QCD is not such a theory.  The perturbative behavior of QCD at high energies is surely relevant to the earliest stages of a heavy-ion collision, and confinement is obviously relevant for late stages.  Our assumption in attempting to compare $\AdS$ calculations with heavy ions is that there is an intermediate regime where QCD is fairly strongly coupled and fairly close to conformal, and that the dynamics of this intermediate regime is crucial to the production of entropy.  Having stressed that these are assumptions which may fail to some degree, we will attempt in this section a comparison of the trapped surface calculation \eno{GotStrapped} to data.  Later on in section~\ref{SLICE} we attempt to make a quantitative estimate of how a failure of these assumptions will affect our results.  

In order to make such a comparison, we must first fix all the parameters appearing on the right hand sides of \eno{GotStrapped}--\eno{E:betatobz}.  We use
\begin{equation}
	L^3/G_5 = 1.9 \,,  \label{LcubedValue}
\end{equation}
which we obtain by comparing the equation of state of the SYM theory to that of lattice QCD, as in \cite{Gubser:2008pc}.  We must also relate $E_\pm$, $z_\pm$, $b$, and $S$ to experimentally measured quantities.  For $E_{\pm}$ we use the total beam energy, which in the case of RHIC collisions reads
\begin{equation}
\label{E:Energydef}
    E_+=E_-= 197\times \frac{\sqrt{s_{NN}}}{2}\,.
\end{equation}
The parameters $z_+$ and $z_-$ give the rms-energy-density-averaged transverse radius of the nuclei. A typical energy distribution for the nucleus is given by a Woods-Saxon profile which has an exponential fall-off. We've fixed $z_{\pm}$ to fit the rms-energy-density-averaged transverse radius of a gold nucleus resulting from a Woods-Saxon profile,
\begin{equation}
\label{E:Goldsize}
 	z_+ = z_- = 4.3\, {\rm fm}\,.
\end{equation}
See \cite{Gubser:2008pc} for details.  Combining \eno{E:betatobz} and~\eno{E:Goldsize} we obtain
 \begin{equation}\label{E:betaAndB}
  \beta \approx 0.12\, {b \over {\rm fm}}
 \end{equation}
for a gold-gold collision.  For $\sqrt{s_{NN}} = 200\,{\rm GeV}$, combining \eno{ZetaPM}, \eno{ZetaDef}, \eno{LcubedValue}, \eno{E:Energydef}, and \eno{E:Goldsize} gives $\zeta \approx 77$.  As remarked in section~\ref{S:Trapped}, this is large enough that just the leading order estimate \eno{GotStrapped} can be used for $b/L$ not too large.  From here on, we will consider only this leading order approximation.

Putting everything together, we can rewrite the entropy bound \eqref{GotStrapped} as  
 \eqn{SIneq}{
  S \geq 35000 \left({\sqrt{s_{NN}} \over 200\, {\rm GeV}} \right)^{2/3} 
    {\sinh^{-1} \beta \over \beta \sqrt{1 + \beta^2}}
 }
for gold-gold collisions.  As explained in the introduction, the amount of entropy produced after a collision of two heavy ions may be inferred from a measurement of the total number of charged particles reaching the detector:
\begin{equation}
\label{E:SandN}
	S \approx 7.5 N_{\rm ch}\,.
\end{equation}
Combining \eno{SIneq} and \eno{E:SandN}, we get
 \eqn{NchargedIneq}{
  N_{\rm charged} \geq 4700 \left({\sqrt{s_{NN}} \over 200\, {\rm GeV}} \right)^{2/3} 
    {\sinh^{-1} \beta \over \beta \sqrt{1 + \beta^2}} \,.
 }
Most of the rest of this section is devoted to confronting the simple formula \eno{NchargedIneq} with data.

\begin{figure}
  \centerline{\includegraphics[width=4in]{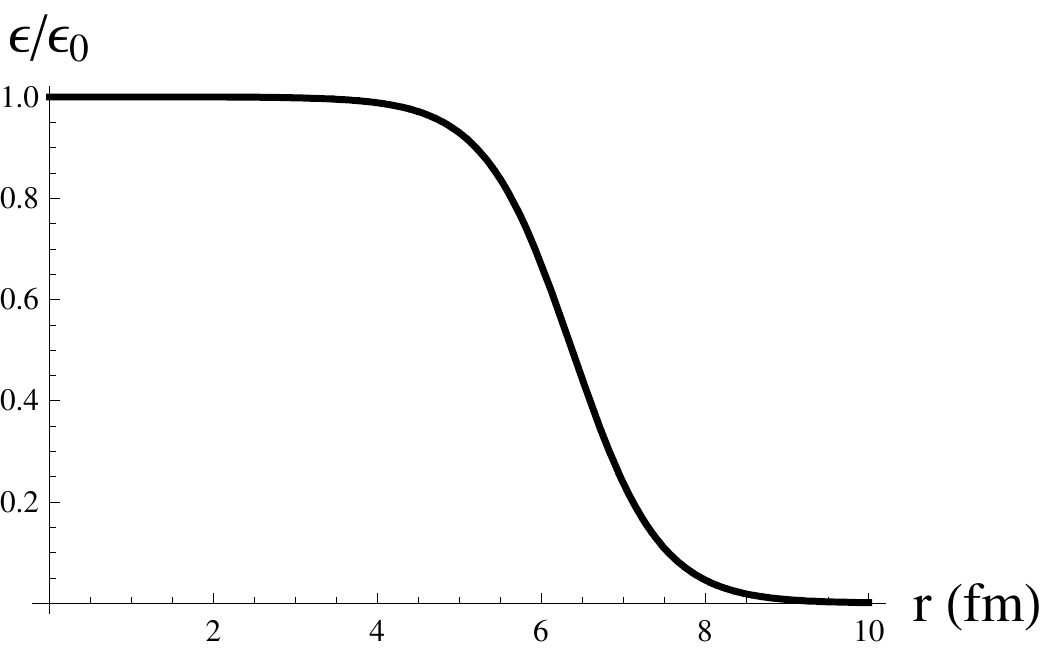}}
  \caption{The energy density of a gold nucleus according to the Woods-Saxon profile, $\epsilon = {\epsilon_0 \over 1 + \exp[(r-R)/a]}$, as a function of the radial distance $r$ to the center of the nucleus.  The energy distribution was normalized so that its value at $r = 0$ is one.  For a gold nucleus, the parameters $R$ and $a$ take the values $\epsilon_0 = 0.159\, {\rm GeV} / {\rm fm}^3$, $R = 6.38\, {\rm fm}$, and $a = 0.535\, {\rm fm}$ \cite{Adams:2004rz}.}
  \label{WoodsSaxon}
\end{figure}

The impact parameter $b$ can be related to the total number of nucleons participating in the collision, $N_{\rm part}$ \cite{Back:2001xy}.  Our description here of how this is done in the Glauber model closely follows \cite{Kharzeev:1996yx,Kharzeev:2000ph,Miller:2007ri}.  Each heavy ion is replaced by a distribution of nucleons, which is proportional to the energy density and is given by a Woods-Saxon profile: see figure~\ref{WoodsSaxon}.  We work with a distribution $\hat\rho$ proportional to energy density but normalized to unity.
The effective distribution in the plane orthogonal to the beam axis (also normalized to unity) is given by
\begin{equation}
	\hat{T}(\vec{r}) = \int \hat\rho(z,\vec{r})dz \,, 
	 \label{TransverseDist}
\end{equation}
where $z$ is the direction of the beam and $\vec{r}$ is a vector in the plane orthogonal to the beam axis.  To determine $N_{\rm part}$, one asks how many nucleons would scatter if there were no other interactions among nucleons other than inelastic nucleon-nucleon scattering, whose cross-section $\sigma_{NN}$ is measured.  In the optical approximation, where one ignores the discreteness of individual nucleons in the distribution \eno{TransverseDist}, the result is
\begin{multline}
\label{E:OpticalGlauber}
	N_{\rm part}(\vec{b}) = N_A \int \hat{T}_A(\vec{r}) \left(
		1- \left(1-\hat{T}_B(\vec{r}-\vec{b})\sigma_{NN}\right)^{N_B}\right)d^2r
		\\{} + N_B \int \hat{T}_B(\vec{r}) \left(
		1- \left(1-\hat{T}_A(\vec{r}-\vec{b})\sigma_{NN}\right)^{N_A}\right)d^2r \,,
\end{multline}
where $N_A$ and $N_B$ are the numbers of nucleons in each nucleus.  We took the values of $\sigma_{NN}$ from  \cite{Kharzeev:2000ph}. A detailed derivation of \eqref{E:OpticalGlauber} can be found in \cite{Miller:2007ri}.

\begin{figure}
  \centerline{\includegraphics[width=5in]{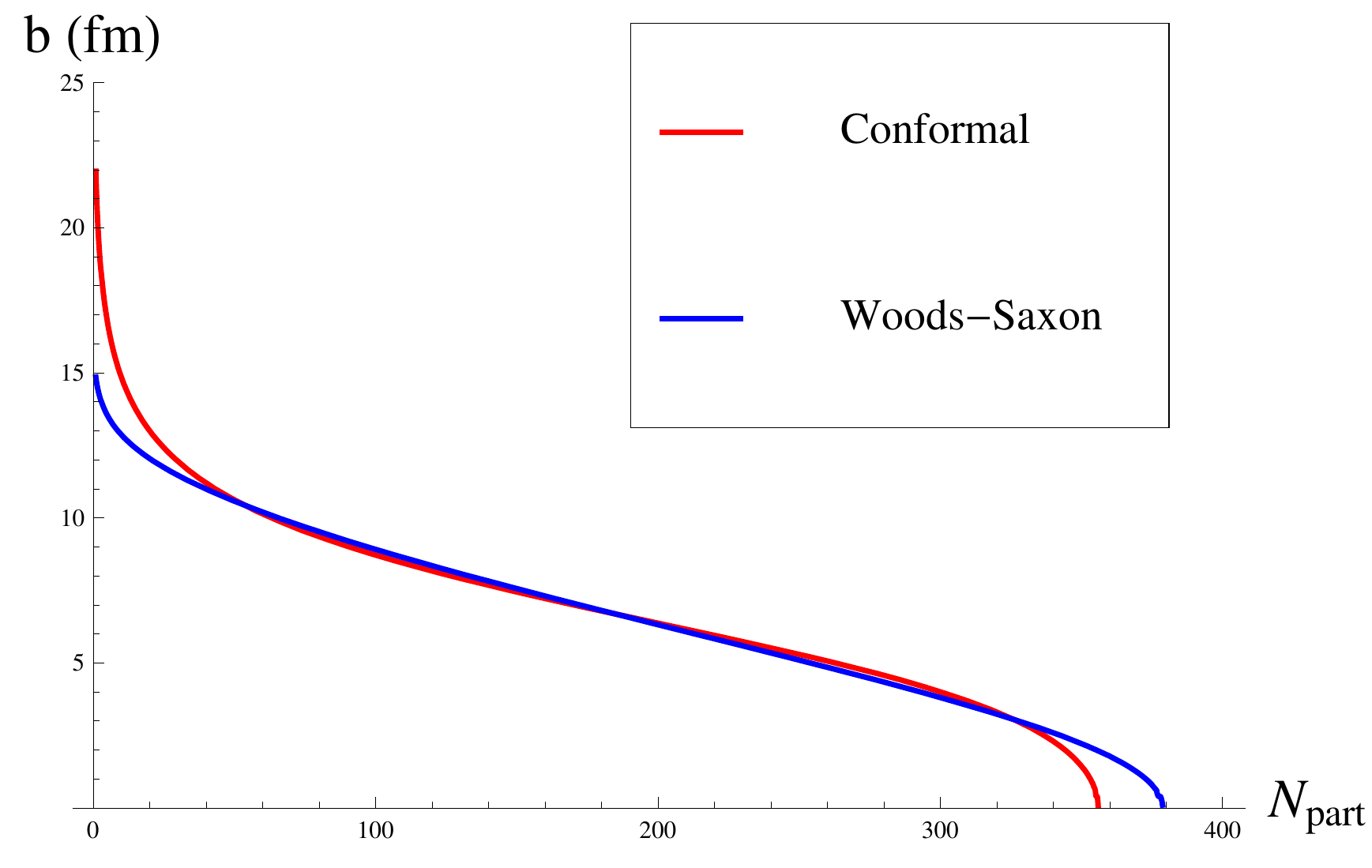}}
\caption{(Color online.)  The impact parameter $b$ as a function of the number of participating nucleons $N_{\rm part}$ in a gold-gold collision, as obtained through optical Glauber calculations at $\sqrt{s_{NN}}=200\,{\rm GeV}$, where $\sigma_{NN}=42\,{\rm mb}$.  The blue curve is based on the standard Woods-Saxon distribution, whereas the red curve is based on the conformal distribution, proportional to \eno{Tmmidentical}. Note that in going from $130\,{\rm GeV}$ to $200\,{\rm GeV}$ the scattering cross section decreases to $41\,{\rm mb}$.\label{F:Glauber}}
\end{figure}

As we have remarked, one usually uses a Woods-Saxon profile to obtain the transverse distribution \eno{TransverseDist} employed in Glauber model calculations.  But the transverse energy distribution dual to a point-sourced shock wave is instead given by \eno{Tmmidentical}, which we will describe as conformal because it preserves an $O(3)$ subgroup of conformal transformations.  It would be more faithful to the $\AdS$ computation to use the conformal distribution instead of Woods-Saxon.  Perhaps surprisingly, for the energy ranges we are considering both these profiles give rather similar results.  See figure~\ref{F:Glauber}.
We therefore employed a Woods-Saxon profile for the Glauber calculations used to compare \eno{NchargedIneq} to data.

\begin{figure}
	\centerline{\includegraphics[width=7in]{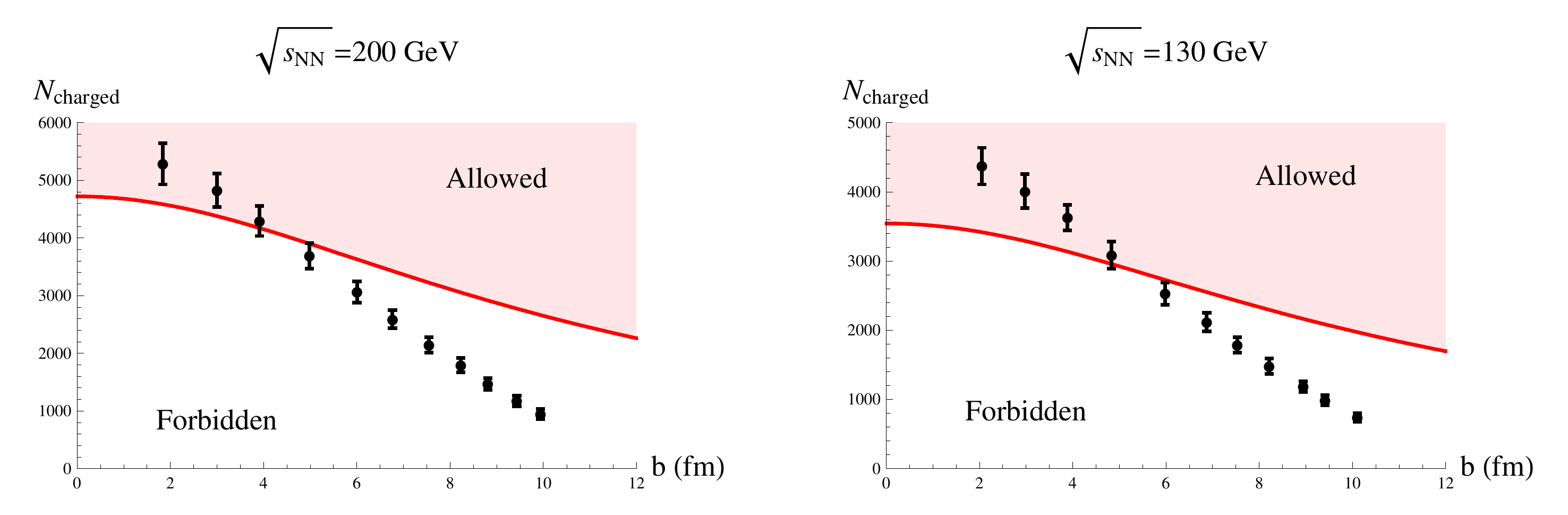}}
\caption{(Color online.)  Total number of charged particles $N_{\rm charged}$ as a function of impact parameter $b$. The data was taken from the PHOBOS experiment \cite{Back:2005hs}.  The red curve corresponds to the lower bound on the number of charged particles which is based on the gauge-gravity duality \eqref{GotStrapped}.}\label{F:impactsimple}
\end{figure}

\begin{figure}
	\centerline{\includegraphics[width=7in]{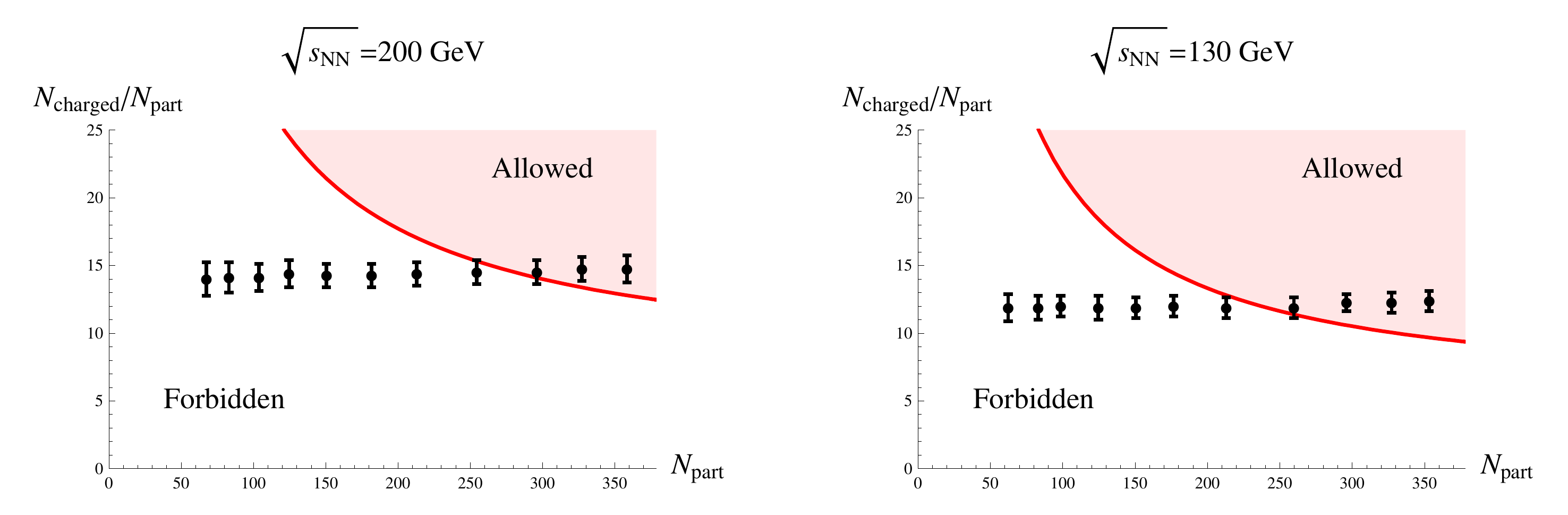}}
\caption{(Color online.)  Plots exhibiting the linear dependence of the total number of charged particles $N_{\rm charged}$ on the number of participating nucleons $N_{\rm part}$. The data was taken from \cite{Back:2005hs}. The shaded red region shows the allowed values of $N_{\rm charged}/N_{\rm part}$, based on \eqref{GotStrapped}.}\label{F:ratiosimple}
\end{figure}

In figure~\ref{F:impactsimple} we compare our lower bound on the entropy as a function of the impact parameter with the PHOBOS data at $\sqrt{s_{NN}}=200$~GeV and $\sqrt{s_{NN}}=130$~GeV, taken from \cite{Back:2005hs} (see also \cite{Nouicer:2002ks}).
Instead of plotting $N_{\rm charged}$ in terms of the impact parameter $b$, it is more common to plot  $N_{\rm charged}/N_{\rm part}$ versus $N_{\rm part}$. We have done so in figure \ref{F:ratiosimple}.

\begin{figure}
	\centerline{\includegraphics[width=7in]{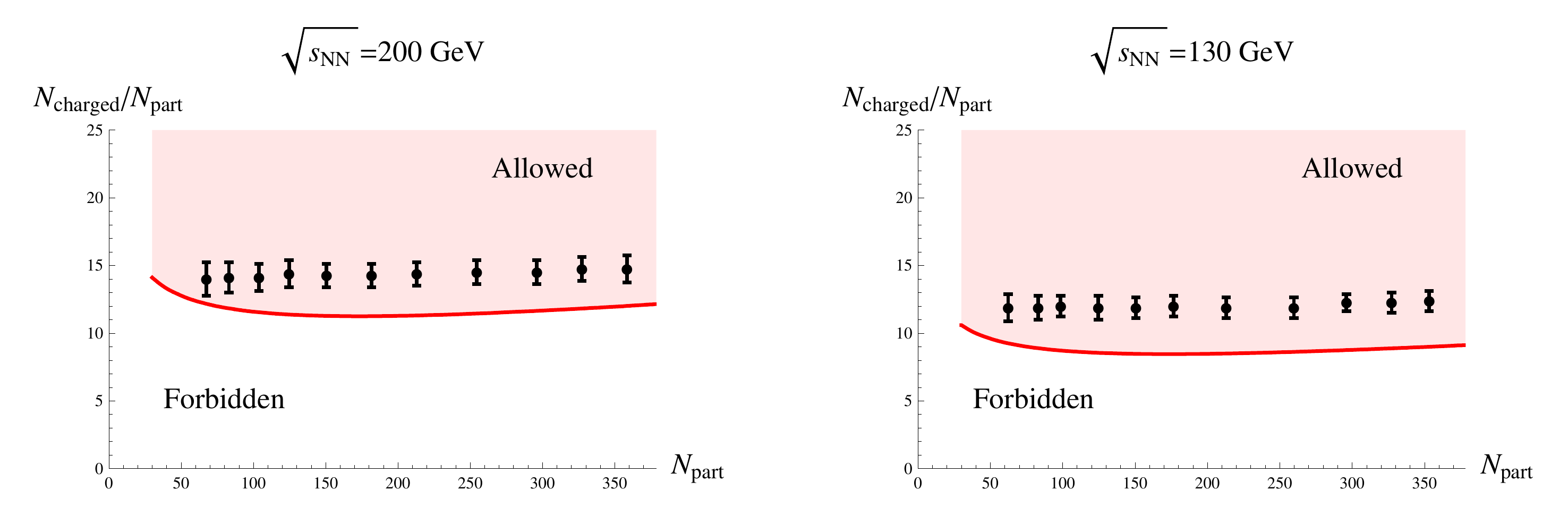}}
\caption{(Color online.)  Plots exhibiting the linear dependence of the total number of charged particles $N_{\rm charged}$ on the number of participating nucleons $N_{\rm part}$. The data was taken from \cite{Back:2005hs}. The shaded red region shows the allowed values of $N_{\rm charged}/N_{\rm part}$, based on rescaling the right hand side \eno{NchargedIneq} by $\left( {N_{\rm part}(b) \over 2 \times 197} \right)^{2/3}$ as explained in the text.
}\label{F:rescaled}
\end{figure}

The mismatch in $b$ dependence between string theory results and the data is due to our not handling infrared effects correctly.  The problem is that in a conformal theory, widely separated objects interact more strongly than they do in QCD.  When two heavy-ions collide in the real world, the nucleons which are not in the collision region (spectators) are not expected to interact. On the other hand, in the collisions we have discussed, energy distributions which are far apart during the time of the collision will produce entropy.  We can mimic the effect of spectators by setting $E_+$ and $E_-$ in \eqref{E:Energydef} equal to the fraction of the energy of each nucleus that participates in the collision according to a Glauber analysis.  In other words, we rescale the energy of a collision with impact parameter $b$ by ${N_{\rm part}(b) \over 2 \times 197}$, where $N_{\rm part}(b)$ is computed via the Glauber model.  As a result, the right hand side of \eno{NchargedIneq} is multiplied by $\left( {N_{\rm part}(b) \over 2 \times 197} \right)^{2/3}$.  The results of such a reinterpretation of the collision energy are shown in figure \ref{F:rescaled}.

While the fit exhibited in figure \ref{F:rescaled} is in good agreement with experiment, the approach on which it is based
is somewhat ad hoc. A more consistent approach to fitting a holographic prediction with the data would be to repeat our calculation in a holographic dual of a confining gauge theory. 

\section{Slicing trapped surfaces}
\label{SLICE}

In the previous section we saw that the absence of confinement in our holographic model leads to a spectator problem in off-center collisions: too much energy far from the center of mass becomes thermalized.  We solved this problem by redefining the energy of the shocks as the energy participating in the collision.  We would prefer a solution where total energy can be cleanly identified between QCD and the five-dimensional bulk, and where the important non-conformal aspects of QCD are captured holographically.  As a first step in this direction, we 
discard the parts of the trapped surface which lie outside a ``safe'' region
 \eqn{E:Safe}{
  z_{\rm IR} > z > z_{\rm UV} \,.
 }
Our rationale for excluding the region $z < z_{\rm UV}$ is that this region corresponds to the ultraviolet physics of QCD, which is asymptotically free.  Partons are nearly free at high energy scales, so they often pass by one another without producing significant energy through scattering.  Of course, there are hard scattering events, and these will be very interesting in LHC heavy-ion collisions.  We are essentially ignoring such events, or folding them into the determination of the parameter $z_{\rm UV}$.

We exclude the region $z > z_{\rm IR}$ because \AdS gets cut off in the infrared by confinement.  The part of the trapped surface that has $z > z_{\rm IR}$ is not relevant to QCD because it relates to processes with energy lower than the confinement scale.  Entropy production in real QCD must happen to some extent at energy scales below confinement, but it is presumably a small effect compared to the entropy production in the deconfined phase provided the total energy of the collision is large enough.

In order to choose sensible values for $z_{\rm IR}$ and $z_{\rm UV}$, we first recall that the ${\rm AdS}_5$-Schwarzschild solution takes the form
 \eqn{AdSSch}{
  ds^2 = {L^2 \over z^2} \left( -h dt^2 + d\vec{x}^2 +
    {dz^2 \over h} \right) \,,
 }
where $h = 1-z^4/z_H^4$ and $L$ is the radius of \AdS.  The temperature is $T = 1/\pi z_H$.  Based on this last relation, we associate physics at a scale $\Lambda$ with the region of \AdS with
 \eqn{zUVDef}{
  z \equiv {1 \over \pi\Lambda} \,.
 }
For the calculations in this section, we choose the values of $z_{\rm UV}$ and $z_{\rm IR}$ that correspond through \eno{zUVDef} to $\Lambda_{\rm UV} = 2\,{\rm GeV}$ and $\Lambda_{\rm IR} = 0.2\,{\rm GeV}$.

In summary: We start with the trapped surfaces obtained in section~\ref{S:Trapped} and intersect them with the ``safe'' region \eno{E:Safe} to obtain a reduced entropy that excludes contributions from the ultraviolet, where we do not trust supergravity, and the infrared, where the bulk geometry should be cut off by effects dual to confinement.

In the case of head-on collisions of objects of equal sizes, finding the area of the trapped surface inside the safe region \eqref{E:Safe} is fairly straightforward due to the $O(2)$ symmetry of the trapped surface in the $x^1x^2$-plane.  Collisions which are not head on and/or those between objects of different sizes, are more challenging to handle.  In the rest of this section we set $z_+ = z_-$, corresponding to equal-sized objects, although it should be possible to generalize our computations to unequal $z_+$ and $z_-$.  We will work at an arbitrary impact parameter $b$.  We find it convenient to switch to a different coordinate system on $H_3$ defined through
 \eqn{hvxvyCoords}{
  X^{-1} &= L \sqrt{1 + v_x^2 + v_y^2} \cosh h \cr
  X^1 &= L \sqrt{1 + v_x^2 + v_y^2} \sinh h \cr
  X^2 &= L v_x \cr
  X^4 &= L v_y \,.
 }
These coordinates have previously been used in \cite{Gawedzki:1991yu}.  When $h$, $v_x$, and $v_y$ are unrestricted, they cover all of $H_3$ in a single coordinate patch.  In fact, the mapping between ${\bf R}^{3}$ and $H_3$ given in \eqref{hvxvyCoords} is, up to a global rescaling, a volume-preserving diffeomorphism, as can be checked by noting that in the $(h, v_x, v_y)$ coordinates the $H_3$ metric,
 \eqn{VolPresH3}{
  ds_{H_3}^2 = {L^2 \over 1 + v_x^2 + v_y^2}
    \bigg[(1 + v_y^2) dv_x^2
    - 2 v_x v_y dv_x dv_y + (1 + v_x^2) dv_y^2 \bigg]
    + L^2 (1 + v_x^2 + v_y^2) dh^2 \,,
 }
has determinant equal to $L^6$.  The $O(2)$ symmetry of the trapped surface manifests itself as rotations in the $X^2X^4$-plane, or as rotations in the $v_x v_y$-plane. A constant $h$ section of the trapped surface is a disk in the $v_x v_y$-plane. Keeping only the leading term in \eqref{E:rhofinal} and using the coordinate transformations \eqref{E:H3coords} and \eqref{hvxvyCoords}, we find that the radius of this disk, $v(h)$, is given by
 \eqn{GotRh}{
  v(h)^2 = 2 {1 + \beta^2 + \zeta^2 \over 1 + 2 \beta^2 + \cosh(2 h)} - 1 \,.
 }
The area of the trapped surface, without cutoffs, is given by
 \begin{equation}
  A_{\rm trapped} = 2 \int_{-\infty}^{\infty} dh \, \pi v(h)^2 dh
   \,. 
 \end{equation}

\begin{figure}
  \centerline{\includegraphics[width=6.5in]{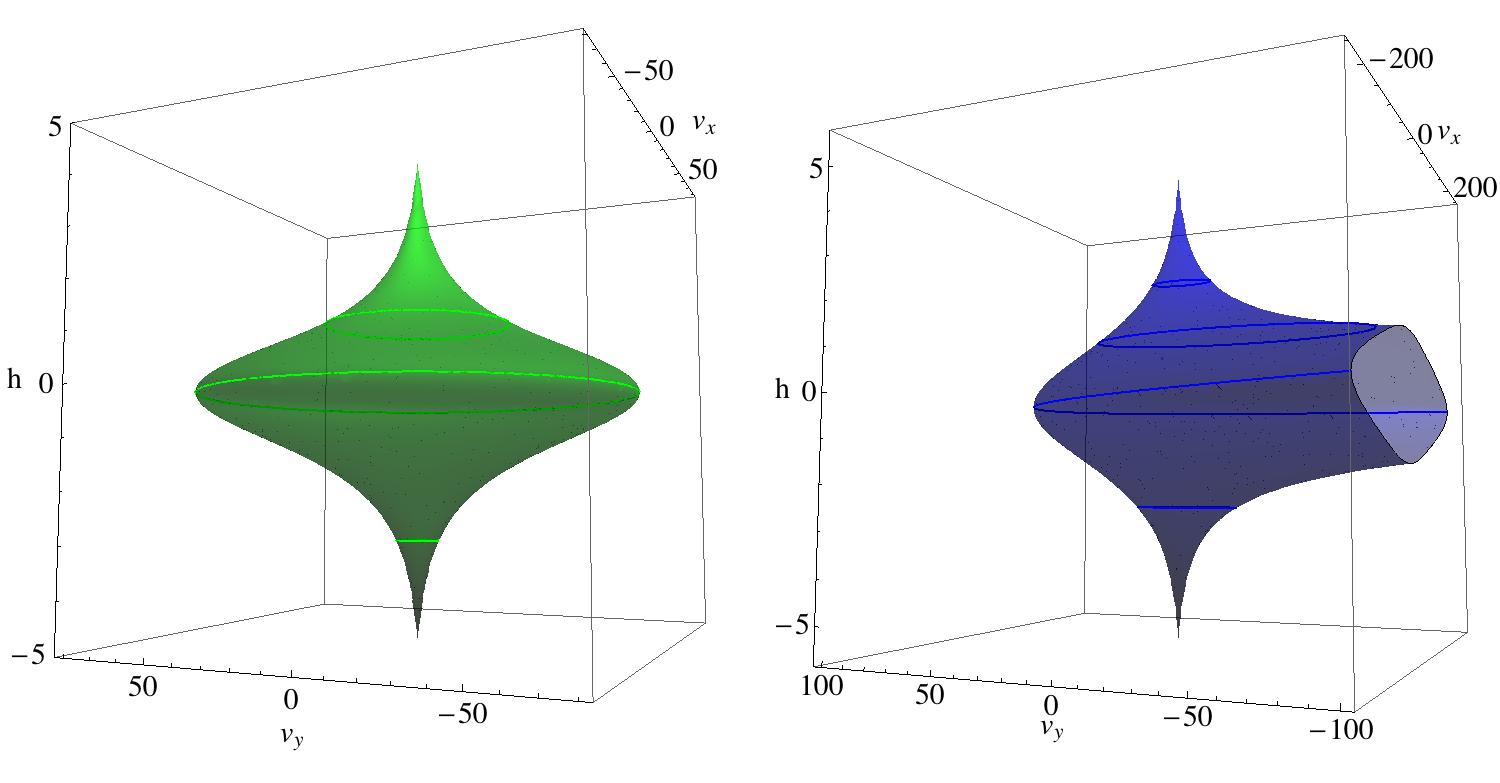}}
  \caption{(Color online.)  A plot of a trapped surface (left) in the $(h,v_x,v_y)$ coordinates for a head-on collision. 
The bright green circles correspond to surfaces of constant $h$. The blue surface (right) corresponds to a surface of constant $z$, whose constant $h$ slices are ellipses shown in dark blue.  At $h=0$, the corresponding ellipse degenerates into a parabola.}\label{F:RBfigs}
 \end{figure}

To implement the UV and IR cutoffs of \eqref{E:Safe} we first need to find out what a surface of constant $z$ looks like in the $(h,v_x,v_y)$ coordinate system.
Using \eqref{E:GlobalToPoincare} and \eqref{hvxvyCoords} one can show that this surface is given by
 \eqn{ConstantzConic}{
    \frac{(v_x-v_x^{\rm center})^2}{a^2}+
      \frac{(v_y-v_y^{\rm center})^2}{b^2}=1\,,
 }
with
\begin{align*}
      (v_x^{\rm center}, v_y^{\rm center}) &= (0, -\frac{L}{z} (\csch h)^2) \\
      a^2&=\frac{L^2}{z^2} (\csch h)^2-1 \\
      b^2&=(\coth h)^2 \left[\frac{L^2}{z^2}(\csch h)^2-1\right]
\end{align*}
for non-vanishing $h$.
Thus, a constant $h\neq 0$ slice of a constant $z$ surface is an ellipse whose eccentricity is
 \eqn[c]{EllipseParams}{
  e = {1\over \cosh h} \,.
 }
When $h=0$ the ellipse degenerates into a parabola symmetric about the $v_y$-axis, as can be seen from the fact that $e = 1$ and $a = \infty$ in this case.\footnote{The parameters characterizing the parabola can be obtained from taking appropriate limits of \eqref{EllipseParams}:  the focal length is given by $\lim_{h \to 0} a (1-e) = {L \over 2 z}$, and the focal point is located at $v_x = 0$ and $v_y = \lim_{h \to 0} v_y^{\rm center} + a e = -{z\over 2 L}$.} See figure \ref{F:RBfigs}.

\begin{figure}
  \centerline{\includegraphics[width=6.5in]{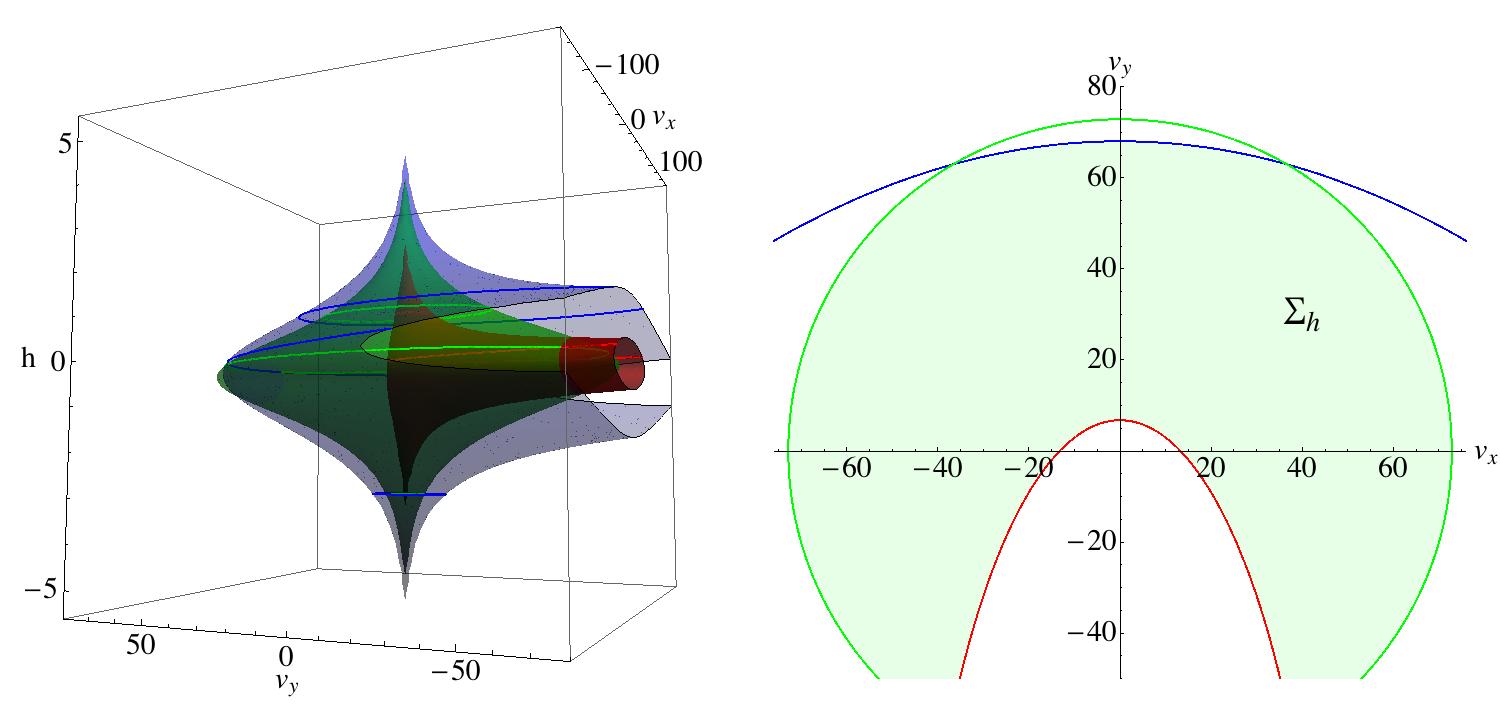}}
  \caption{(color online) Plots of the trapped surface together with the UV and IR regions. On the left we show a plot of the trapped surface (green) in the $(h,v_x,v_y)$ coordinate system. Some lines of constant $h$ are lightly colored for emphasis. The UV surface and the IR surface are colored blue and red, respectively. Lines of constant $h\neq 0$ on these surfaces are ellipses. On the right we've plotted a constant $h$ section of the trapped surface and the UV and IR cutoff surfaces.}\label{hSlice}
 \end{figure}

In figure \ref{hSlice} we've plotted the trapped surface (green) together with the $z=z_{\rm UV}$ surface (blue) and the $z=z_{\rm IR}$ surface (green). Pictorially, we need to compute the volume of the green region which lies between the blue and red surfaces. In practice, it is easiest to compute the area of a constant $h$ section $\Sigma_h$ of the trapped surface restricted to $z_{\rm UV}<z<z_{\rm IR}$ and then integrate over $h$:
\begin{equation}
	\mathcal{A} = \int_{-\infty}^{\infty} {\rm Vol}(\Sigma_h) dh
\end{equation}
where
\begin{equation}
 	\Sigma_h = \{(v_x,v_y):\ (v_x,v_y,h)\in \mathcal{C} \text{ and } z_{\rm UV}<z<z_{\rm IR} \}.
\end{equation}
In figure~\ref{hSlice} we show a typical constant-$h$ slice.

It is straightforward to see that the ellipse in \eqref{ConstantzConic} intersects the circle given in \eqref{GotRh} at the $v_y$ coordinate
 \eqn{vyIntersection}{
  v_y^{\rm int} = {L\over z} - {\sqrt{2(1 +
    \beta^2 + \zeta^2)} \cosh h \over
    \sqrt{1 + 2 \beta^2 + \cosh 2 h} } \,.
 }
The $v_x$ coordinate of the intersection can be computed from either \eqref{ConstantzConic} or \eqref{GotRh}, but the resulting formula will not be needed.
Let's denote by ${\cal A}^{\rm IR}_h$ the two-dimensional area common to the disk of radius \eqref{GotRh} and the region $z>z_{\rm IR}$ (which in figure~\ref{hSlice} corresponds to the intersection between the disk and the interior of the inner ellipse) and by ${\cal A}^{\rm UV}_h$ the area common to the same disk and the region $z>z_{\rm UV}$ (which in figure~\ref{hSlice} corresponds to the intersection between the disk and the interior of the outer ellipse).  The total area of the trapped surface is then
 \eqn{SboundCut}{
  A_{\rm slice} = 2 \left( 
    \int_{-\infty}^\infty dh\, {\cal A}^{\rm UV}_h - 
    \int_{-\infty}^\infty dh\, {\cal A}^{\rm IR}_h \right) \,.
 }
The quantities ${\cal A}^{\rm IR}_h$ and ${\cal A}^{\rm UV}_h$ can be computed analytically for every value of $h$.  If, for example, the inner (IR) ellipse in figure~\ref{hSlice} is contained completely inside the disk, then ${\cal A}^{\rm IR}_h$ should just equal the area of the ellipse;  if the disk is contained inside the ellipse, then ${\cal A}^{\rm IR}_h$ should equal the area of the disk;  if the inner ellipse intersects the boundary of the disk (as drawn in figure~\ref{hSlice}), one needs to add the part of the area of the ellipse above the horizontal line passing through the intersection point in figure~\ref{hSlice} to the part of the area of the disk below this intersection point.  The explicit expression for $A_{\rm slice}$ is not very illuminating, so we won't reproduce it here.

\begin{figure}
   \centerline{\includegraphics[width=5in]{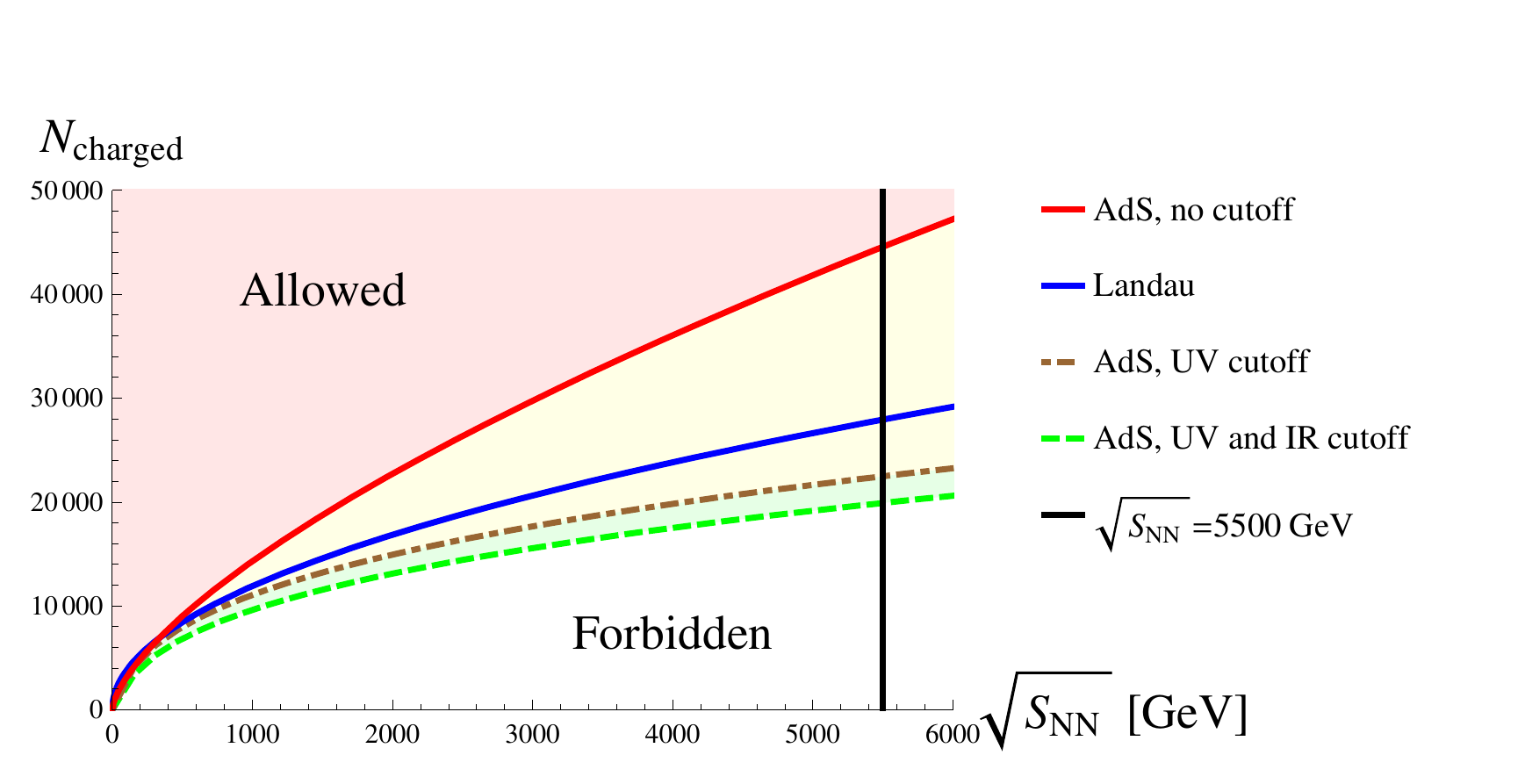}}
   \caption{Total number of charged particles $N_{\rm charged}$ for central Pb-Pb collisions over a range of energies. The red curve represents the $b = 0$ limit of the AdS prediction \eqref{NchargedIneq}.  The blue curve represents the prediction of the Landau model (see, for example, section~2.3 of \cite{Gubser:2008pc}).  The brown dot-dashed curve represents the AdS prediction \eqref{SboundCut} with a UV cutoff at $\Lambda_{\rm UV} = 2\, {\rm GeV}$.  The blue dashed curve represents the AdS prediction \eqref{SboundCut} with a UV cutoff at $\Lambda_{\rm UV} = 2\, {\rm GeV}$ and an IR cutoff at $\Lambda_{\rm IR} = 0.2\, {\rm GeV}$.  The black vertical line marks the value of $\sqrt{s_{NN}}$ expected to be attained at the LHC\@.}\label{CentralSlice}
 \end{figure}

What we are interested in is how $S_{\rm slice} \equiv A_{\rm slice} / 4G_5$ compares with the entropy $S_{\rm trapped}$ computed from the entire trapped surface.  For central collisions, the results are shown in figure~\ref{CentralSlice}, converted to $N_{\rm charged}$ using the formula \eno{SN}.\footnote{It should be noted that using \eno{SN} over the large range of energy shown in figure~\ref{CentralSlice} is not necessarily justified.  Investigations along the lines of \cite{Pal:2003rz} based on LHC data should help clarify the relation between $N_{\rm charged}$ and $S$ at higher energies.}  The UV cutoff doesn't significantly decrease the predicted bound on $N_{\rm charged}$ at RHIC energies, but as one proceeds to large $E$, its effect is dramatic: whereas $S_{\rm trapped} \propto E^{2/3}$, we find $S_{\rm slice} \propto E^{1/3}$ in the limit of large $E$.  The IR cutoff decreases the predicted bound on $N_{\rm charged}$, but not by very much.  To obtain the curves plotted in figure~\ref{CentralSlice}, we used $L^3/G_5 = 1.9$, $z_* = 4.4\, {\rm fm}$, and $E_{\rm beam} = 208 {\sqrt{s_{NN}} \over 2}$, as appropriate for Pb.

\begin{figure}
	\centerline{\includegraphics[width=7in]{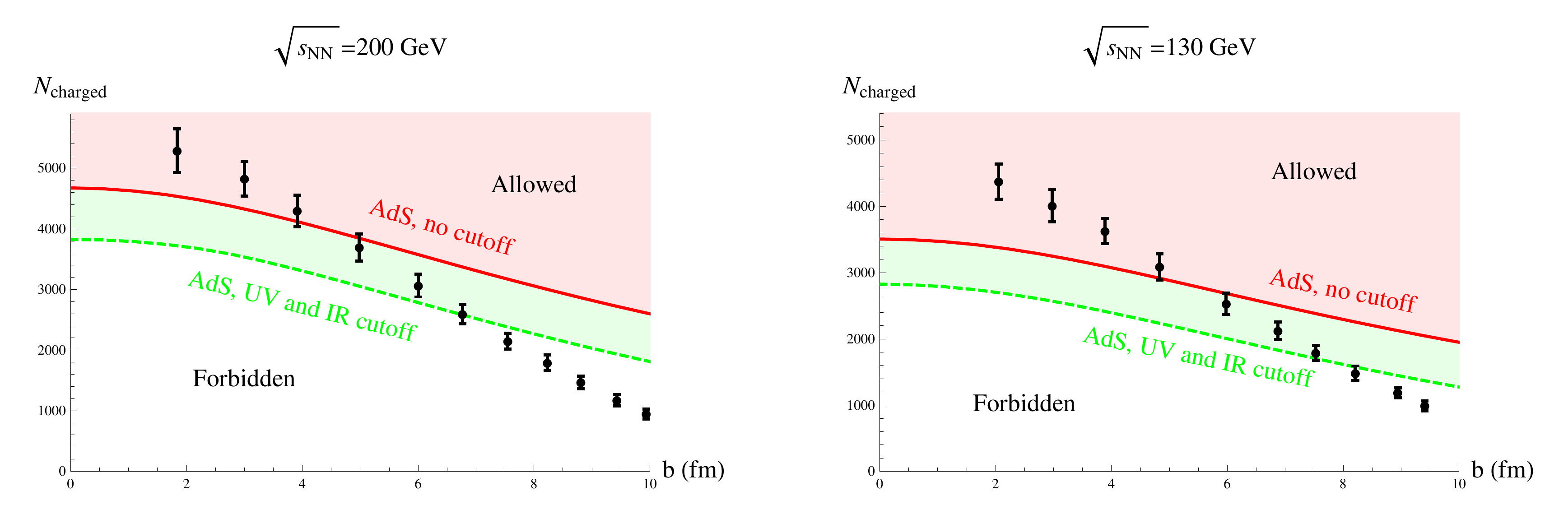}}
\caption{Total number of charged particles $N_{\rm charged}$ as a function of impact parameter $b$. The data was taken from the PHOBOS experiment \cite{Back:2001xy,Nouicer:2002ks}. The red curve corresponds to the lower bound on the number of charged particles which is based on the the dual of a slice of AdS  \eqref{GotStrapped}.}\label{F:impactSlice}
\end{figure}

\begin{figure}
	\centerline{\includegraphics[width=7in]{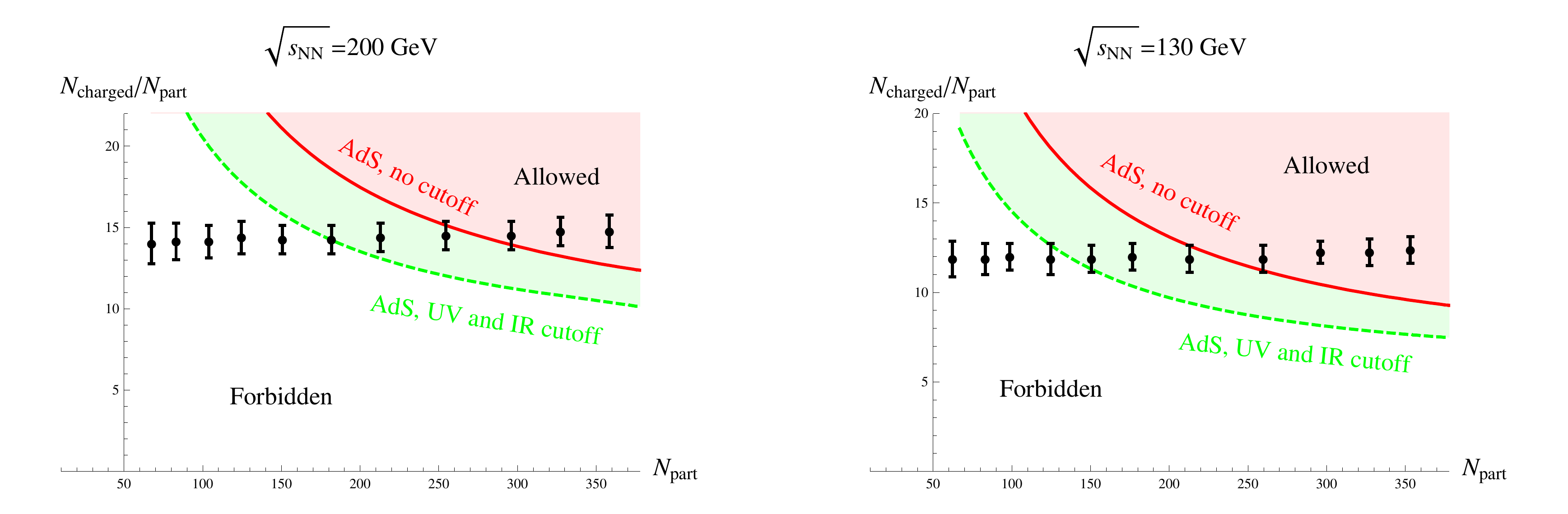}}
\caption{Plots exhibiting the linear dependence of the total number of charged particles $N_{\rm charged}$ on the number of participating nucleons $N_{\rm part}$. The data was taken from \cite{Back:2001xy,Nouicer:2002ks}. The shaded red region shows the allowed values of $N_{\rm charged}/N_{\rm part}$ computed via \eqref{SboundCut}.}\label{F:ratioSlice}
\end{figure}

Now let us turn to the case of non-zero impact parameter.  In figures~\ref{F:impactSlice} and~\ref{F:ratioSlice} we plotted the lower bound on $N_{\rm charged}$ corresponding to $S \geq S_{\rm slice}$ for gold-gold collisions at $\sqrt{s_{NN}} = 200\, {\rm GeV}$ and $\sqrt{s_{NN}} = 130\, {\rm GeV}$ together with the data from the PHOBOS experiment \cite{Back:2001xy,Nouicer:2002ks}.  The values for $L^3/G_5$, $z_\pm$, and $E$ used to make these plots are the ones given in the previous section.
Figures~\ref{F:impactSlice} and~\ref{F:ratioSlice} are an improvement relative to the fit in figures \ref{F:impactsimple} and \ref{F:ratiosimple} in the sense that more data points are in the allowed region.  Also, the string theory predictions at larger $b$ are lowered by an incrementally larger factor than for small $b$.  However, the string theory predictions still show too much entropy produced in glancing collisions as compared to central collisions. 
Comparing figure \ref{F:impactSlice} to figure \ref{F:rescaled}, we conclude that
our slicing procedure does not capture the effects of confinement as well as rescaling the energy did. It may be that a better treatment of the UV and IR cutoffs---in other words, better control over the non-conformal behavior of the theory---will yield a better justified fit to the data.

\section{Discussion}
\label{DISCUSSION}

Our main formal results, summarized in \eno{E:rhofinal} and~\eno{StrappedBetter}, are approximate analytic expressions for the size and shape of a trapped surface produced in off-center collisions of point-sourced shock waves in \AdS, in a limit where the trapped surface is much bigger than the radius of curvature of \AdS.  The precise limit is $\zeta \gg 1$ and $\zeta \gg \beta$.  It is quite striking that analytic results can be obtained, given that in flat space, the computation of trapped surfaces is a tricky numerical problem: see for example \cite{Yoshino:2002tx}.

When using this formal result to make direct estimates of total multiplicities in heavy-ion collisions, we need to convert various AdS${}_5$ quantities to equivalent QCD observables.  If we interpret the energy of the colliding shocks as the total beam energy, then our results are in disagreement with the data at impact parameters greater than $4$ to $5$ fm.  
But if, instead, we interpret the energy of the colliding shocks as the energy participating in the collision as calculated via the Glauber model, then, as shown in figure \ref{F:rescaled}, agreement with data is good.

Another challenge to our AdS-based model is the scaling $S \propto E^{2/3}$ at high energies: the power $2/3$ is probably too large. We've tried to address this problem by slicing off parts of AdS space. This crude approach is motivated by the idea that entropy production comes predominantly from processes whose energy scale is above the confinement scale, but not so far above it as to be substantially suppressed by asymptotic freedom.  Independent of the extent to which the slicing approach is justified, the calculations of section~\ref{SLICE} provide information about where in the transverse $H_3$ space (the slice of \AdS below the transverse $x^1x^2$-plane) the entropy comes from.

To gain some intuition about the results of the slicing calculations, let's consider a head-on collision with $z_+ = z_- = L$, $E_+ = E_- = E$, and $\zeta = (2 E G_5 / L^2)^{1/3}$, in accord with \eno{ZetaPM}.  First let's consider the shape of the trapped surface in the absence of any cutoffs.  Let $z_{\rm min}$ and $z_{\rm max}$ be the minimum and maximum values of $z$ on the trapped surface, and let $x_{\perp, \rm max}$ be the maximum value of $\sqrt{(x^1)^2 + (x^2)^2}$ on the trapped surface.  Straightforward calculations to leading order in large $\zeta$ yield
 \eqn{zxValues}{
  {z_{\rm min} \over L} = {1 \over 2\zeta} \qquad
   {z_{\rm max} \over L} = 2\zeta \qquad
   {x_{\perp,\rm max} \over L} = \zeta \,.
 }
If we compare to a central gold-gold collision at $\sqrt{s_{NN}} = 200\,{\rm GeV}$ by setting $L = 4.3\,{\rm fm}$ and $L^3 / G_5 = 1.9$, then $\zeta = 77$.  (Note that setting $L^3 / G_5 = 1.9$ makes $\zeta$ quite close to $(EL)^{1/3}$.)  Combining the result \eno{zxValues} for $z_{\rm min}$ with the conversion formula $\Lambda = 1/\pi z$ between depth and energy scale, one finds
 \eqn{LambdaValue}{
  \Lambda_{\rm max} \equiv {1 \over \pi z_{\rm min}} \approx
   2.2 \, {\rm GeV} \,.
 }
In comparing to QCD processes, it is tempting to regard $\Lambda_{\rm max}$ as the maximum temperature of the fireball produced in a heavy-ion collision.  But the trapped surface is far from being an equilibrated black hole horizon, so $\Lambda_{\rm max}$ probably shouldn't be regarded as the peak temperature.   Instead, it is the maximum energy scale of processes in the fireball that dominate entropy production.  On this interpretation, the result \eno{LambdaValue} doesn't seem to us unreasonable in comparison with QCD\@.  What seems more troublesome is the result for $x_{\perp,\rm max}$ in \eno{zxValues}: plugging in $L = 4.3\,{\rm fm}$ and $\zeta = 77$ gives
 \eqn{xPerpMax}{
  x_{\perp,\rm max} = 330\,{\rm fm} \,,
 }
which, as noted in \cite{Lin:2009pn}, seems way too high for a heavy-ion collision.

Let's take a closer look at the distribution of the entropy in the transverse plane.  For a central collision, it is fairly straightforward to calculate the fraction $f(x_\perp)$ of the entropy that lies within the region
 \eqn{CylinderRegion}{
  (x^1)^2 + (x^2)^2 \leq x_\perp^2 \,.
 }
It is interesting to note that although $f(x_\perp)$ is non-zero all the way out to $x_{\perp,\rm max} = \zeta L$, its second moment is much smaller:
 \eqn{rmsXPerp}{
  x_{\perp,\rm rms} \equiv \sqrt{
    \int_0^{x_{\perp,\rm max}} dx_\perp \, 
       x_\perp^2 f'(x_\perp)}
    \approx L \sqrt{2 \log 2\zeta - 3}
    \approx L \sqrt{{2 \over 3} \log {8EL} - 3}\,,
 }
where in the last expression we recalled that $\zeta \approx (EL)^{1/3}$ when $L^3/G_5 = 1.9$.  We explain how to obtain \eno{rmsXPerp} in Appendix~\ref{SIZE}.  With parameters as described above, $x_{\perp,\rm rms}$ is roughly $2.7 L \approx 11\,{\rm fm}$.  Recalling that $L$ is the energy-weighted rms radius of each shock wave, we see from \eno{rmsXPerp} that the entropy-weighted rms radius $x_{\perp,\rm rms}$ is larger only by a modest factor, even at very large $\zeta$.  As we explain in Appendix~\ref{SIZE}, for $x_\perp / L$ of order unity, the distribution of entropy in the transverse plane as quantified by $f(x_\perp)$ is not so distant from sensible expectations for QCD.  The surprising result \eno{xPerpMax}, then, is not disastrous: it indicates that the tail of the entropy distribution is too long, not that the dominant part of the entropy distribution is unreasonable.

Although $f(x_\perp)$ has a precise meaning on the gravity side of the duality, it is probably only approximately correct to translate it into the fraction of the entropy in the boundary theory within a radius $x_\perp$ of the collision point.  Indeed, all results describing the position of the trapped surface in \AdS should be interpreted with caution when passing to the field theory side of the gauge-string duality.  The trapped surfaces we have constructed stretch from $t=-\infty$ to $t=0$, but it would be obviously at odds with causality to say that the entropy is produced before the collision.  In general, a black hole horizon is not something one can describe locally in time.  It has to do with whether particles in a region of spacetime can escape to some asymptotic infinity.  Trapped surfaces also have a non-local character: although the differential equation \eno{hpmEOM} is local, the matching conditions \eno{E:Dagain1}--\eno{E:Dagain2} are not.  The trapped surfaces we have constructed identify a region of spacetime from which a test particle cannot emerge once it enters---modulo some conjectural points relating to Cosmic Censorship.  In particular, the test particle can be separated in the transverse $(x^1,x^2)$ coordinates by as much as $x_{\perp,\rm max}$ and still fall inside the trapped surface.  If we use \eno{xPerpMax}, what this says is that a test particle more than $300\,{\rm fm}$ away from the colliding nuclei can eventually thermalize with the resulting fireball.  Thermalization of such a distant test particle is impossible in QCD because of the short-range nature of the interactions. 
This underscores once again the real differences between real-world QCD and a strongly coupled conformal field theory.

In Appendix~\ref{SIZE}, we show that the IR cutoff lowers $x_{\perp, \rm max}$ from $330\, {\rm fm}$ to $13.8\, {\rm fm}$ in a central collision with the choice of parameters indicated after \eno{zxValues}.  This dramatic reduction of $x_{\perp,\rm max}$ is associated with less than a $20\%$ reduction in the total entropy.  In addition, $x_{\perp, \rm rms}$ is reduced by the cutoffs from $11\,{\rm fm}$ to roughly $5\,{\rm fm}$ for the parameters described just after \eno{zxValues}.

In summary: The difficulty we had matching our results to total multiplicity of off-center collisions indicates that the trapped surface in the dual of a strongly coupled conformal field theory has too great a tendency to swallow up regions of spacetime that are far from the collision.  In the conformal field theory, the fireball produced in an off-center collision has too great a tendency to thermalize with energy far from the collision point.  Such energy corresponds to the spectators in an off-center heavy-ion collision.  We suspect that this tendency goes hand in hand with excessively long tails of the entropy distribution in central collisions.  Both difficulties would probably be cured in a holographic theory with a mass gap.

\section*{Acknowledgments}
We thank B.~Cole, M.~Gyulassy, S.~Lin, H.~Nastase, J.~Maldacena, J.~Noronha, E.~Shuryak, P.~Steinberg, and B.~Zajc for useful discussions.  This work was supported in part by the Department of Energy under Grant No.~DE-FG02-91ER40671 and by the NSF under award number PHY-0652782.

\clearpage

\begin{appendix}
\section{Boundary conditions on $h_\pm(\rho, \theta)$ at small $\rho$}
\label{MATCH}

In this section we explain why we didn't include terms of the form $D_i/\rho^3$ in \eqref{hSoln}.  The starting point is the equation satisfied by $h_+$, which for simplicity we'll denote by $h$:
 \eqn{hEQ}{
  \left( \nabla^2_{H_3} - {3\over L^2} \right) h = 0 \,.
 }
Parameterizing $H_3$ by $(r, \theta, \phi)$ so that the metric is given by \eqref{E:lineH3r}, the most general solution to \eqref{hEQ} that is invariant under additive shifts in $\phi$ are
 \eqn{hSolnR}{
  h &= \sum_{\ell\geq 0} \Bigg[ U^{(\ell)} P_\ell (\cos \theta) \left({r\over L} \right)^\ell
    {}_2 F_1 \left({\ell - 1\over 2}, {\ell + 3 \over 2}, \ell + {3 \over 2}, -{r^2 \over L^2} \right)\cr
    &+ V^{(\ell)} P_\ell (\cos \theta) \left({L\over r} \right)^{\ell + 1}
    {}_2 F_1 \left(-{\ell\over 2} - 1, 1 - {\ell \over 2}, {1\over 2} - \ell , -{r^2 \over L^2} \right) \Bigg]\,,
 }
where $P_\ell(x)$ are the Legendre polynomials, ${}_2 F_1$ is the hypergeometric function, and $U^{(\ell)}$ and $V^{(\ell)}$ are integration constants.  Since the hypergeometric functions are regular at $r = 0$, it follows that the $U$-solution is regular at $r = 0$, while the $V$-solution is not.  The large $r$ behavior is a bit more complicated:
 \eqn{hLargeR}{
  h &= \sum_{\ell\geq 2} \Bigg[ U^{(\ell)} P_\ell (\cos\theta)
    \left( a_{-1}^{(\ell)} {r\over L} +
    a_1^{(\ell)} {L\over r} + {\cal O}(\log r/r^3) \right)
    + V^{(\ell)} P_\ell (\cos\theta)
    \left( b_3^{(\ell)} {L^3\over r^3} + {\cal O}(1/r^5) \right) \Bigg] \cr
    &{}+ \left[ \left(U^{(1)} - 2 V^{(1)} \right) {r\over L} + {3 \over 4} V^{(1)} {L^3 \over r^3} + {\cal O}(1/r^5)\right] \cr
    &{}+ \left[ \left(U^{(0)} + 2 V^{(0)} \right) {r\over L} + {1\over 2} \left(U^{(0)} + 2 V^{(0)} \right) {L \over r}
    - {1\over 8} U^{(0)} {L^3 \over r^3} + {\cal O}(1/r^5) \right] \,,
 }
where
 \eqn{abDef}{
  a_{-1}^{(\ell)} &= {2^{1-\ell} \Gamma\left({1\over 2}
    - \ell\right) \over \sqrt{\pi} \Gamma(2 - \ell)} \cr
  a_1^{(\ell)} &= {(\ell + 2) \Gamma\left({1\over 2} - \ell\right)
    \over 2^{\ell + 1} \sqrt{\pi} \Gamma(1 - \ell)}  \cr
  b_3^{(\ell)} &= {(-1)^\ell \Gamma\left({1\over 2} - \ell \right)
    \Gamma(\ell + 3) \over 2^{\ell + 3} \sqrt{\pi} } \,.
 }

As explained in section \eqref{S:Trapped}, equation \eqref{hEQ} can also be solved in a coordinate system where $H_3$ is parameterized by $(\rho, \theta, \phi)$ and the metric is \eqref{zetaMet}.  The parameter $\zeta \equiv r / \rho L$ appearing in \eqref{zetaMet} is a large expansion parameter.  By expanding
 \eqn{hSeries}{
  h(\rho, \theta) = h_0(\rho, \theta)
    + {1 \over \zeta^2} h_2(\rho, \theta) + \cdots \,,
 }
one obtains\footnote{Actually, the next order term in \eqref{hSeries} is of order $\zeta^{-4} \log \zeta$, so not all terms in this expansion take the form \eqref{hSoln}.}
 \begin{subequations}\label{hSoln}
 \begin{align}
  h_0 &= C_0(\theta) \rho + D_0(\theta) {1\over \rho^3} \label{hZeroSolnAgain}  \\
  h_2 &= C_2(\theta) \rho + D_2(\theta) {1\over \rho^3} +
    {1 \over 4\rho} \left( \partial_\theta^2 +
    \cot\theta \partial_\theta + 2 \right) C_0(\theta) \cr
    &{}-{1\over 12 \rho^5} (\partial_\theta^2 +
    \cot \theta \partial_\theta + 6) D_0 (\theta)\,.
    \label{hTwoSolnAgain}
 \end{align}
 \end{subequations}

The matched asymptotic expansion technique implies that the small $\rho$ limit of \eqref{hSoln} needs to be matched onto the large $r$ limit of \eqref{hSolnR} (equation \eqref{hLargeR}) after plugging in $r = \rho L \zeta$ in \eqref{hLargeR}.  To do so, we first write
 \eqn{CDHarmonics}{
  C_{2k} = \sum_{\ell \geq 0} C_{2k}^{(\ell)} P_\ell (\cos\theta) \qquad
  D_{2k} = \sum_{\ell \geq 0} D_{2k}^{(\ell)} P_\ell (\cos\theta) \,.
 }
We find
 \eqn{Match}{
  U^{(\ell)} &= {a_{-1}\over \zeta} \left(C_0^{(\ell)}
    + {1\over \zeta^2} C_2^{(\ell)} + \cdots \right)
      \qquad \text{for $\ell\geq 2$} \cr
  U^{(1)} &= {1\over \zeta} \left( C_0^{(1)}
    + {1\over \zeta^2} C_2^{(1)} + \cdots \right)
    + {8 \zeta^3 \over 3} \left( D_0^{(1)}
    + {1\over \zeta^2} D_2^{(1)} + \cdots \right) \cr
  U^{(0)} &= - 8 \zeta^3 \left( D_0^{(0)}
    + {1\over \zeta^2} D_2^{(0)} + \cdots \right) \cr
  V^{(\ell)} &= b_3 \zeta^3 \left(D_0^{(\ell)}
    + {1\over \zeta^2} D_2^{(\ell)} + \cdots \right)
      \qquad \text{for $\ell\geq 2$} \cr
  V^{(1)} &= {4 \zeta^3 \over 3} \left( D_0^{(1)}
    + {1\over \zeta^2} D_2^{(1)} + \cdots \right) \cr
  V^{(0)} &= {1\over 2 \zeta} \left( C_0^{(0)}
    + {1\over \zeta^2} C_2^{(0)} + \cdots \right)
    + 4 \zeta^3 \left( D_0^{(0)}
    + {1\over \zeta^2} D_2^{(0)} + \cdots \right) \,.
 }
One can arrive at the first and fourth relations above by comparing the $a_{-1}^{(\ell)}$ and $b_3^{(\ell)}$ terms in \eqref{hLargeR} to \eqref{hZeroSolnAgain} and the first two terms in \eqref{hTwoSolnAgain}.  The $a_1^{(\ell)}$ term in \eqref{hLargeR} is the same as the third term in \eqref{hTwoSolnAgain} to leading order, but this is just a consistency check.  A similar consistency check consists of comparing the fourth term in \eqref{hTwoSolnAgain} to the ${\cal O}(1/r^5)$ term that multiplies $V^{(\ell)}$ which was not written explicitly in \eqref{hLargeR}.  The other relations in \eqref{Match} can be found by making similar comparisons of the terms in the second and third lines of \eqref{hLargeR} to \eqref{hSoln}.

The main application of this matching procedure is that if one requires regularity of the solution at $r = 0$, one necessarily has $D_0^{(\ell)} = D_2^{(\ell)} = 0$ for all $\ell$.  To see this, note that regularity implies $V^{(\ell)} = 0$.  From the fourth and fifth relations in \eqref{Match}, this immediately implies that $D_0^{(\ell)} = D_2^{(\ell)} = 0$ for $\ell \geq 1$.  The same is true for $\ell = 0$ since the $D_0^{(0)}$ and $D_2^{(0)}$ terms in the last equation in \eqref{Match} dominate over the $C_{2k}^{(0)}$ terms at large $\zeta$.  Note that from the form of the formulas in \eqref{Match} one may infer that $D_{2k}^{(\ell)} = 0$ for all $\ell \geq 1$ and all $k$, but it is not clear, for example, that $D_4^{(0)}$ should vanish.

\section{The transverse size of the trapped surface}
\label{SIZE}

In this appendix we consider the computation of the fraction $f(x_\perp)$ of the area of the trapped surface formed in a head-on collision that falls within the region $(x^1)^2 + (x^2)^2 \leq x_\perp^2$.  For simplicity we set $z_+ = z_- = L$ and $E_+ = E_- = E$.  Throughout, we work to leading order in large $\zeta = (2 E G_5/L^2)^{1/3}$.

The trapped surface has the familiar $O(3)$ symmetry that preserves the point of impact on the transverse space $H_3$.  The region $(x^1)^2 + (x^2)^2 \leq x_\perp^2$ preserves an $O(2)$ subgroup of this symmetry.  As in section~\ref{SLICE}, it is advantageous to use $(h,v_x,v_y)$ coordinates in the presence of a residual $O(2)$ symmetry.  But because the particular $O(2)$ we are now considering is  a different subgroup of $O(3)$ from the one preserved in off-center calculations, we use a different identification from \eno{hvxvyCoords}:
 \eqn{NewHV}{
  X^{-1} &= L \sqrt{1 + v_x^2 + v_y^2} \cosh h = 
    {z \over 2} \left( 1 + {L^2 + (x^1)^2 + (x^2)^2 \over z^2} 
      \right) \cr
  X^1 &= L v_x = L {x^1 \over z}  \cr
  X^2 &= L v_y = L {x^2 \over z}  \cr
  X^4 &= -L \sqrt{1 + v_x^2 + v_y^2} \sinh h = 
    {z \over 2} \left( -1 + {L^2 - (x^1)^2 - (x^2)^2 \over z^2}
      \right) \,.
 }
The trapped surface covers the region of $H_3$ with
 \eqn{TrappedIneq}{
  (X^1)^2 + (X^2)^2 + (X^4)^2 \leq \zeta^2 L^2 \,,
 }
to leading order in $\zeta$.  Denoting $v = \sqrt{v_x^2 + v_y^2}$, we see that \eno{TrappedIneq} may be recast as
 \eqn{TrappedAgain}{
  v \leq v(h) \equiv \sqrt{\zeta^2 \sech^2 h - \tanh^2 h} \,,
 }
where $h$ must be restricted to the range $-h_0 \leq h \leq h_0$, with
 \eqn{hZeroDef}{
  h_0 = \sinh^{-1} \zeta \approx \log 2\zeta \,.
 }
Next, we observe that
 \eqn{xot}{
  {(x^1)^2 + (x^2)^2 \over L^2} = 
    {(X^1)^2 + (X^2)^2 \over (X^{-1}+X^4)^2} = 
   e^{2h} \left( 1 - {1 \over 1+v_x^2+v_y^2} \right) \,.
 }
This implies that the inequality $(x^1)^2+(x^2)^2 \leq x_\perp^2$ is equivalent to
 \eqn{xpIneq}{
  v \leq {x_\perp / L \over 
    \sqrt{e^{2h} - x_\perp^2/L^2}} \qquad\hbox{or}\qquad
   h \leq \log {x_\perp \over L} \,.
 }
Because the volume form on $H_3$ is just $L^3 dh \wedge dv_x \wedge dv_y$, the fraction $f(x_\perp)$ can be computed as the ratio of the volume in the $(h,v_x,v_y)$ coordinate space satisfying both \eno{TrappedAgain} and~\eno{xpIneq} to the volume satisfying only \eno{TrappedAgain}.  At large $\zeta$, one finds
 \eqn{fLargeZeta}{
  f'(x_\perp) \approx {2 L^2 x_\perp \over (L^2 + x_\perp^2)^2} 
    \sqrt{1 - {x_\perp^2 \over \zeta^2 L^2}} \,.
 }
From \eno{fLargeZeta} it follows that
 \eqn{RMSlog}{
  {1 \over L^2} 
   \int_0^{x_{\perp,\rm max}} dx_\perp \, x_\perp^2 f'(x_\perp)   
     \approx 2 \log 2\zeta - 3 \,,
 }
which gives the result quoted in \eqref{rmsXPerp}.  It is worth noting that $f'(x_\perp)$ falls off as $1/x_\perp^3$ but is cut off at $x_{\perp,\rm max} = \zeta L$ by the square root that appears in \eqref{fLargeZeta}.

To go further, let us define the distribution of entropy over the transverse plane as
 \eqn{StrappedDist}{
  {dS_{\rm trapped} \over dx^1 dx^2} = {S_{\rm trapped} \over
    2\pi x_\perp} f'(x_\perp) \,.
 }
Using \eqref{fLargeZeta}, we find that at large $\zeta$,
 \eqn{logDist}{
  {1 \over S_{\rm trapped}} {dS_{\rm trapped} \over dx^1 dx^2} 
    \propto {1 \over (L^2 + x_\perp^2)^2} 
     \sqrt{1- {x_\perp^2 \over \zeta^2 L^2}} \,.
 }
If we assume that 
 \eqn{dSsat}{
  {d^2 S_{\rm trapped} \over dx^1 dx^2} \sim Q_s^2 \,,
 }
where $Q_s$ is the saturation scale,\footnote{The saturation scale is a characteristic scale of transverse momenta of gluons in the early stages of a highly relativistic collision of nuclei.} then we find
 \eqn{QsDist}{
  Q_s(x_\perp) \propto {\zeta \over L^2 + x_\perp^2} 
    \left({1- {x_\perp^2\over \zeta^2 L^2}}\right)^{1/4}\,,
 }
which again holds only in the limit of large $\zeta$.  This formula shows that the saturation scale has approximately a $1/(L^2 + x_\perp^2)$ dependence which is cut off at $x_\perp = x_{\perp, \rm max} = \zeta L$, where $Q_s(x_{\perp, \rm max}) = 0$.  It is striking that essentially the same spatial dependence, $Q_s \propto 1/(R^2 + x_\perp^2)$, has been advocated from the perspective of the color glass condensate \cite{Iancu:2007st}.  A difference is that  the parameter $R$ of \cite{Iancu:2007st} is expected to grow logarithmically with energy, due to spread of the saturating region into the outer corona, whereas in our case, $L$ is simply a constant.  The slow growth of the rms size \eno{rmsXPerp} of the trapped surface is due mostly to the lengthening of the power-law tail.  In summary: transverse distributions reveal a qualitative, or even semi-quantitative, similarity between trapped surfaces in \AdS and the CGC approach; but the reasons for slow growth of transverse size with energy differ.

In the CGC approach, the saturation scale at $x_\perp = 0$ is expected to grow as a slow power: $Q_s^2 \propto E^\lambda$ where $\lambda \approx 0.28$ (see for example \cite{Kharzeev:2001yq}).  Using \eno{QsDist} and the scaling $\zeta \sim E^{1/3}$, we find instead $Q_s^2 \propto E^{2/3}$ up to logarithmic corrections.  This discrepancy probably owes to the fact that the QCD scaling $Q_s \propto E^\lambda$ is essentially a perturbative result, and hence distant from the strong coupling regime in which we work.  Clearly, this is closely related to the expectation in CGC treatments that multiplicities will scale as a smaller power of $E$ than the $E^{2/3}$ that we find.  Thus we hold out some hope that an appropriate treatment of the ultraviolet physics, beyond the supergravity approximation, might lead to a more systematically accurate account of saturation physics as well as total multiplicity in a holographic framework.

Including IR and UV cutoffs changes somewhat the story presented above.  A surface of constant $z$ is described by 
 \eqn{ConstantzSurface}{
  v = \sqrt{ {e^{2 h} L^2 \over z^2} - 1} \,,
 }
and a careful account of the intersections of the IR and UV cutoffs with the trapped surface and the $x_\perp$ cutoff \eqref{xpIneq} yields
 \eqn{xpMaxUVIRCutoffs}{
  x_{\perp, {\rm \max}} \approx \sqrt{2 L z_{\rm IR} \zeta} 
    - {L^2 + z_{\rm IR}^2 \over 2 \sqrt{2 L z_{\rm IR} \zeta}}
 }
in the limit of large $\zeta$.  Our preferred choice of $z_{\rm IR}$, corresponding to $\Lambda_{\rm IR} = 0.2\, {\rm GeV}$, reduces the value of $x_{\perp, {\rm max}}$ from $330\, {\rm fm}$ in absence of cutoffs to $13.8\, {\rm fm}$ with cutoffs.
\begin{figure}
 \begin{center}
  \includegraphics[width=3in]{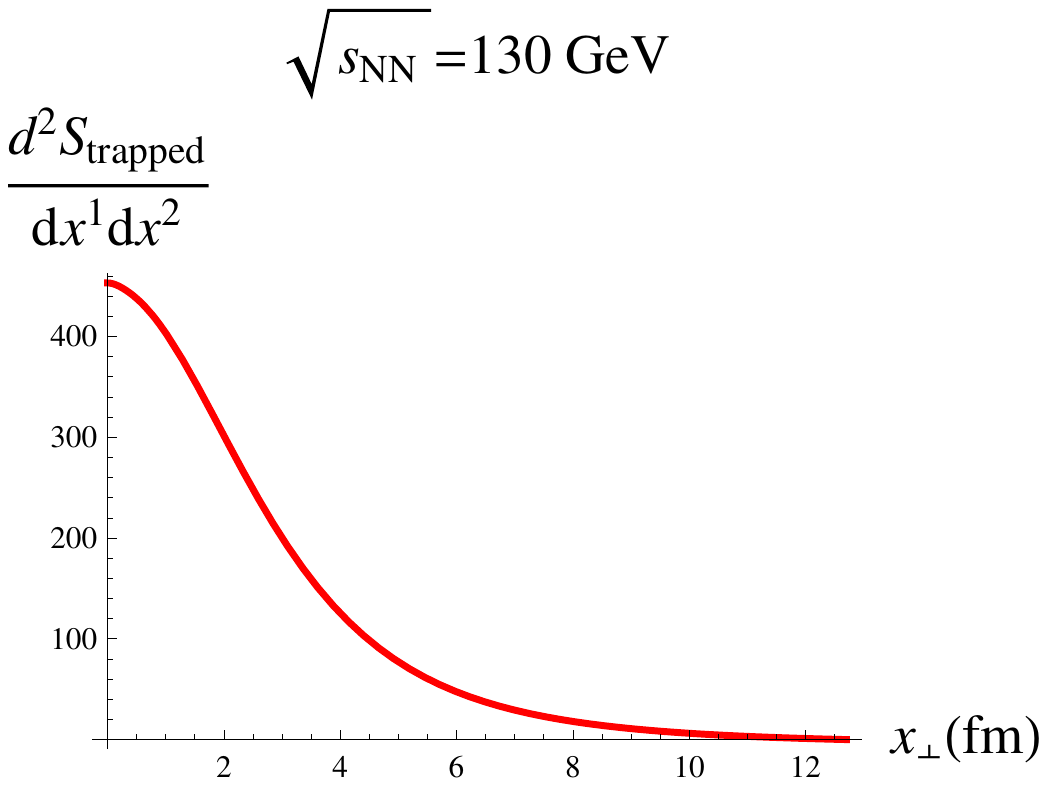}
	\includegraphics[width=3in]{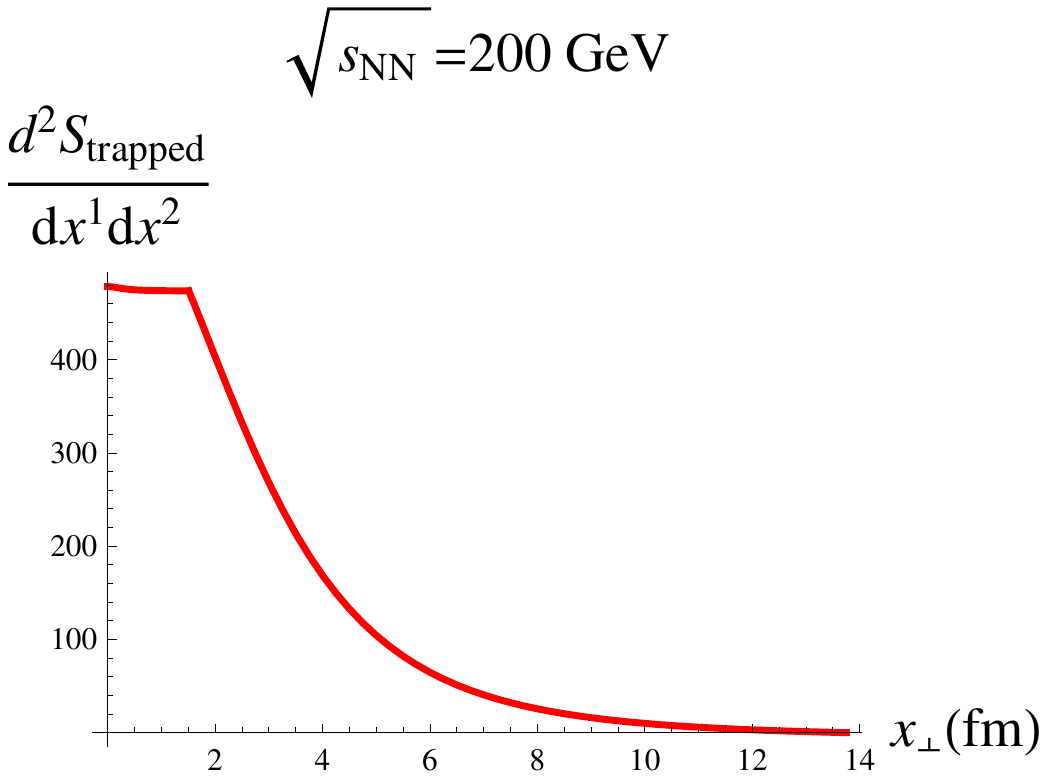} \\[15pt]
  \includegraphics[width=3in]{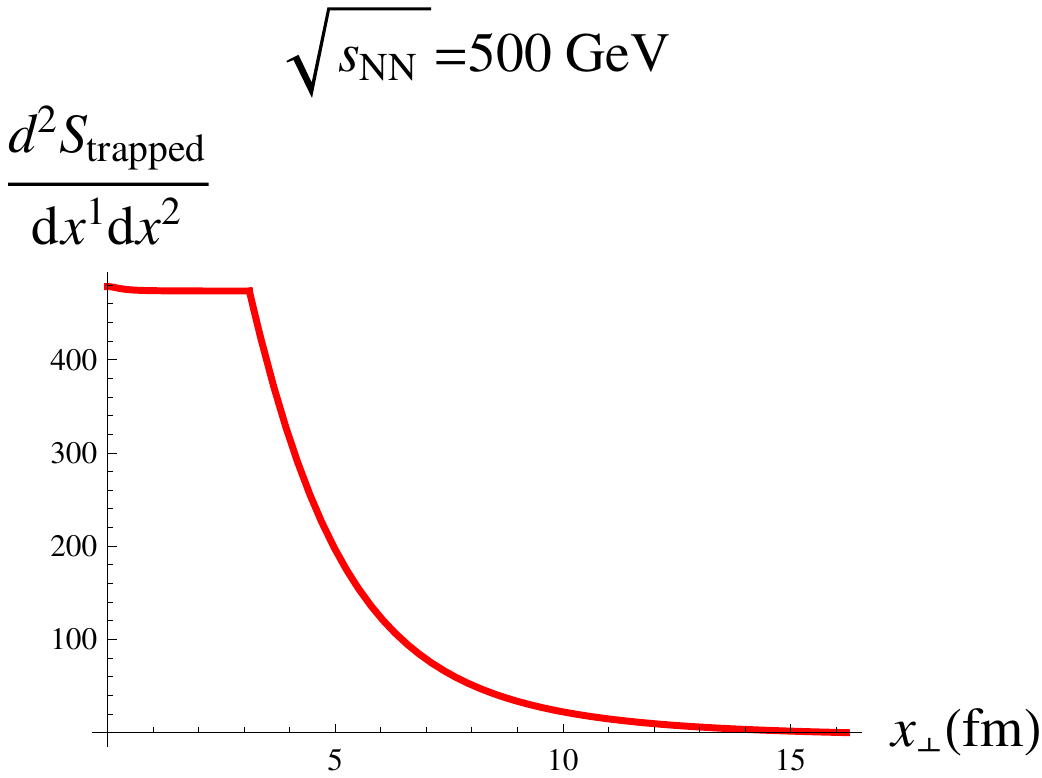}
 \end{center}
\caption{The distribution of the trapped surface entropy in the transverse plane for $\sqrt{s_{NN}} = 130\, {\rm GeV}$, $200\, {\rm GeV}$, and $500\, {\rm GeV}$, in the presence of UV and IR cutoffs.  We used $L^3/G_5 = 1.9$, $L = 4.3\, {\rm fm}$, $\Lambda_{\rm IR} = 0.2\, {\rm GeV}$, and $\Lambda_{\rm UV} = 2\, {\rm GeV}$, as appropriate for a comparison to gold-gold collisions.}\label{EntropyDensityPlots}
\end{figure}
In figure~\ref{EntropyDensityPlots} we show the lower bound ${d^2 S_{\rm trapped} \over dx^1 dx^2}$ for the entropy density in the case of gold-gold collisions at $\sqrt{s_{NN}} = 130\, {\rm GeV}$, $200\, {\rm GeV}$, and $500\, {\rm GeV}$.  In making these plots we set $L^3 / G_5 = 1.9$, $L = 4.3\, {\rm fm}$, $\Lambda_{\rm IR} = 0.2\, {\rm GeV}$, and $\Lambda_{\rm UV} = 2\, {\rm GeV}$ as discussed in sections~\ref{SYMPheno} and~\ref{SLICE}.  The IR and UV cutoffs alter somewhat the scaling ${d^2 S_{\rm trapped} \over dx^1 dx^2} \sim 1/(L^2 + x_\perp^2)^2$ at the values of $\sqrt{s_{NN}}$ considered above, but this scaling becomes more and more accurate in the intermediate $x_\perp$ region at larger $\zeta$.  At $x_\perp = 0$, ${d^2 S_{\rm trapped} \over dx^1 dx^2}$ approaches a $\zeta$-independent constant in the limit of large $\zeta$.  Assuming again that the entropy density is proportional to $Q_s^2$, one can see that this scaling is significantly different from the result $Q_s \sim \zeta \sim E^{1/3}$ in absence of cutoffs, and from the CGC scaling $Q_s \sim E^{0.14}$.  What this shows is that changing the way we deal with UV physics does indeed affect significantly the scaling of $Q_s$ with energy at $x_\perp = 0$.

Our approach of entirely ignoring the ultraviolet region of $\AdS$ (the region where $z<z_{\rm UV}$) is probably too abrupt.  Better would be to have some understanding of how entropy production weakens gradually as one departs from the regime of validity of supergravity.  Such an understanding seems far out of reach starting from first principles in string theory.  It may be that CGC results on the scaling of $Q_s$ provide some hints about the correct stringy dynamics of the ultraviolet region of the bulk in the true holographic dual of QCD.

\end{appendix}

\bibliographystyle{ssg}
\bibliography{shocks}

\end{document}